\documentclass[12pt]{article} \baselineskip=.2cm \textwidth=165mm \textheight=22cm \voffset
-1.5cm \hoffset -1cm

\usepackage{amssymb}
\usepackage{amsfonts}
\usepackage{latexsym}
\usepackage{amsthm}

\usepackage[english]{babel}

\newtheorem{theorem}{\bf Theorem}[section]
\newtheorem{lemma}[theorem]{\bf Lemma}
\newtheorem{corollary}[theorem]{\bf Corollary}
\newtheorem{proposition}[theorem]{\bf Proposition}
\def\remark{\vskip 4pt \noindent {\it Remark.}\ }

\renewcommand{\theequation}{\arabic{section}.\arabic{equation}}

\begin{document}


\def\l{\lambda}
\def\p{\psi}   \def\r{\rho}
\def\s{\sigma}
\def\ve{\varepsilon}
\def\vt{\vartheta}
\def\vp{\varphi}
\def\t{\tau}

\def\cA{{\cal A}}
\def\bH{{\bf H}}
\def\cH{{\cal H}}
\def\cS{{\cal S}}

\def\el2{\ell^{\,2}}
\def\dlt{{\scriptstyle \triangle}}

\let\le\leqslant \let\ge\geqslant

\def\R{{\mathbb R}} \def\C{{\mathbb C}}
\def\D{{\mathbb D}} \def\T{{\mathbb T}}
\def\N{{\mathbb N}}

\def\Re{\mathop{\rm Re}\nolimits}
\def\Im{\mathop{\rm Im}\nolimits}
\def\Ran{\mathop{\rm Ran}\nolimits}
\def\res{\mathop{\rm res}\limits}
\def\Iso{\mathop{\rm Iso}\nolimits}

\newcommand{\oo}[1]{{\mathop{#1}\limits^{\,\circ}}\vphantom{#1}}
\newcommand{\po}[1]{{\mathop{#1}\limits^{\phantom{\circ}}}\vphantom{#1}}
\newcommand{\nh}[1]{{\mathop{#1}\limits^{{}_{\,\bf{\wedge}}}}\vphantom{#1}}
\newcommand{\nc}[1]{{\mathop{#1}\limits^{{}_{\,\bf{\vee}}}}\vphantom{#1}}
\newcommand{\nt}[1]{{\mathop{#1}\limits^{{}_{\,\bf{\sim}}}}\vphantom{#1}}


\title {The inverse problem for perturbed\\ harmonic oscillator on the half-line}
\author{
Dmitry Chelkak\begin{footnote}{Correspondence author. Dept. of Math. Analysis, Math. Mech.
Faculty, St.Petersburg State University. Universitetskij pr. 28, Staryj Petergof, 198504
St.Petersburg, Russia. Partly supported by grants VNP Minobrazovaniya 3.1--4733, RFFR
03--01--00377 and NSh--2266.2003.1. E-mail: delta4@math.spbu.ru}
\end{footnote}\ \ and\ \ Evgeny Korotyaev\begin{footnote}{
Institut f\"ur  Mathematik,  Humboldt Universit\"at zu Berlin. Rudower Chaussee 25, 12489
Berlin, Germany. Partly supported by DFG project BR691/23-1. E-mail:
evgeny@math.hu-berlin.de}\end{footnote}} \maketitle

\begin{abstract}
\noindent We consider the perturbed harmonic oscillator $T_D\psi=-\psi''+x^2\psi+q(x)\psi$,
$\psi(0)=0$ in $L^2(\R_+)$, where $q\in\bH_+=\{q', xq\in L^2(\R_+)\}$ is a real-valued
potential. We prove that the mapping $q\mapsto{\rm spectral\ data}={\rm \{ eigenvalues\ of\
}T_D{\rm \}}\oplus{\rm \{norming\ constants\}}$ is one-to-one and onto. The complete
characterization of the set of spectral data which corresponds to $q\in\bH_+$ is given.
Moreover, we solve the similar inverse problem for the family of boundary conditions
$\psi'(0)=b \psi(0)$, $b\in\R$.
\end{abstract}

\section{Introduction}

Consider the Schr\"odinger operator
$$
{\rm H}=-\frac{\partial^2}{\partial {\bf x}^2}+{\bf x}^2+q(|{\bf x}|),\ \ \ {\bf x}\in \R^3,
$$
acting in the space $L^2(\R^3)$. We set $x=|{\bf x}|\ge0$ and assume $q$ is a real-valued
bounded function. The operator ${\rm H}$ has pure point spectrum. Using the standard
transformation $u({\bf x})\to xu({\bf x})$ and expansion in spherical harmonics, we obtain
that ${\rm H}$ is unitary equivalent to a direct sum of the Schr\"odinger operators acting
on $L^2(\R_+)$. The first operator from this sum is given by
$$
T_D\psi=-\psi''+x^2\psi+q(x)\psi\,,\qquad \psi(0)\!=\!0\,,\qquad x\!\ge\!0\,.
$$
The second is $-\frac{d^2}{dx^2}+x^2+\frac{2}{x^2}+q(x)$ etc. Below we consider the simplest
case, i.e., the operator $T_D$\,. In our paper we assume that
$$
q\in\bH_+=\biggl\{q\!\in\!L^2(\R_+):q',xq\!\in\!L^2(\R_+)\biggr\}\,.
$$
This class of potentials is convenient to solve the inverse problem as it was shown in
\cite{CKK}, which is devoted to the analogous problem on the whole real line. Recall that the
complete characterization of the set of spectral data that corresponds to potentials
$q\!\in\!L^2$ is unknown.

Define the unperturbed operator $T_D^0\psi\!=\!-\psi''\!+\!x^2\psi$\,, $\psi(0)\!=\!0$\,. The
spectrum $\s(T_D)$ of $T_D$ is the increasing sequence of simple eigenvalues
$\lambda_{2n+1}\!=\!\lambda_{2n+1}^0\!+\!o(1)$\,, $n\!\ge\!0$\,, where
$\l_{2n+1}^0\!=\!4n\!+\!3$, $n\!\ge\!0$\,, are the eigenvalues of $T_D^0$\,. The spectrum
$\sigma(T_D)$ does not determine $q(x)$ uniquely, see Theorems 2.2 and 2.3. Then what does the
isospectral set
$$
\Iso_D(q)=\{p\in {\bf H}_+:\, \l_{2n+1}(p)=\l_{2n+1}(q) {\rm\ for\ all\ } n\ge 0\}
$$
of all potentials $p$ with the same Dirichlet spectrum as $q$ look like?

We come to the inverse problem. It has two parts: i) to describe the set $\Iso_D(q)$;
\linebreak ii) to characterize all sequences of real numbers which arise as the Dirichlet
spectra of $q\!\in\!\bH_+$\,. In order to describe $\Iso_D(q)$\,, we need "additional"\
spectral data. Below we consider two choices of such data:

1) The Neumann spectrum $\sigma(T_N)$ of $q(x)$, where $T_N\psi=-\psi''+x^2\psi+q(x)\psi$\,,
$\psi'(0)\!=\!0$\,.  The spectrum $\sigma(T_N)$ is the increasing sequence of simple
eigenvalues $\l_{2n}\!=\!\lambda_{2n}^0\!+\!o(1)$\,, where $\l_{2n}^0=4n\!+\!1$\,,
$n\!\ge\!0$\,. The corresponding inverse spectral problem  $q\mapsto
\{\lambda_{2n}\}_{n=0}^{\infty}\cup\{\lambda_{2n+1}\}_{n=0}^{\infty}$ is equivalent to the
inverse spectral problem on the whole real line with even potentials $q(x)\!=\!q(-x)$\,,
$x\!\in\!\R$ (see Theorem \ref{*H+2SpecChar} and Sect. 3).

2) The sequence of the so-called norming constants for eigenfunctions of $T_D$\,.
The corresponding inverse problem is the main point of our paper  (see Theorem \ref{*H+DChar}
and Sect. 4).

We also consider the inverse problem for the family of operators $\{T_b\}_{b\in \R}$ given by
$$
T_b \psi=-\psi''+x^2\psi+q(x)\psi\,,\qquad \psi'(0)\!=\!b \psi(0)\,,\qquad x\!\ge\!0\,,
$$
see Theorem \ref{*H+NChar} and Sect. 5. Note that the analysis of the whole family
$\{T_b\}_{b\in\R}$ is simpler than the case of the fixed constant $b\in \R$. It is similar to
the Sturm-Liouville problem on the unit interval (\cite{IT}, \cite{IMT}, \cite{RT}). We give
the explicit expression of $b$ in terms of spectral data in Theorem \nolinebreak \ref{*KasSD}.

Our approach is a generalization of the method applied in  \cite{CKK} and \cite{PT} (devoted
to the inverse Dirichlet problem on $[0,1]$). There is a big difference (see e.g. \cite{GS})
between the case of the whole real line and the case of the half-line. The main point in the
inverse problem for the perturbed harmonic oscillator $T_D$ on $\R_+$ is the characterization
of "additional" spectral data, i.e. the characterization of the set $\Iso_D(q)$ (see \cite{GS}
about the connectedness of this set in various classes of potentials). We introduce the
standard norming constants $s_{2n+1}(q)$ by (\ref{SnDefD}), the same way as in \cite{CKK},
\cite{PT}. Unfortunately, we have two problems: i) \nolinebreak the sequence
$\{s_{2n+1}(q)\}_{n=0}^{\infty}$ doesn't belong to a "proper"\ Hilbert space; ii) the
eigenvalues and the norming constants are not "independent coordinates"\ in the space of
potentials.

Roughly speaking, the set of all sequences $\{s_{2n+1}(p)\}_{n=0}^{\infty}$\,,
$p\!\in\!\Iso_D(q)$\,, essentially depends on the spectrum $\{\l_{2n+1}(q)\}_{n=0}^\infty$\,.
Recall that in the case of the Sturm-Liouville operators $-y''\!+\!q(x)y$ on $[0,1]$ with the
mixed boundary conditions $y(0)\!=\!y'(1)\!=\!0$\begin{footnote}{More general, $y(0)\!=\!0$\,,
$y'(1)\!=\!ky(1)$ with any fixed $k$.}\end{footnote} (see \cite{K}) the eigenvalues and the
norming constants are not independent since in this case we have the identity similar to
(\ref{KisEqual}). Note that the dependence between $\{\l_{2n+1}(q)\}_{n=0}^\infty$ and
$\{s_{2n+1}(p)\}_{n=0}^{\infty}$ is more complicated. Therefore, we need an essential
modification of the norming constants. We define the new coordinates
$\{r_{2n+1}\}_{n=0}^{\infty}$ by (\ref{RnDefD}). The new parameters
$\{r_{2n+1}\}_{n=0}^{\infty}$ are "independent coordinates"\ and
$\{r_{2n+1}\}_{n=0}^{\infty}\!\in\!\ell_{3/4}^2$ (see Theorem \ref{*H+DChar}). Moreover, the
spectral mapping $q\mapsto \left(\{\mu_{2n+1}(q)\}_{n=0}^\infty\,,q(0)\,,
\{r_{2n+1}(q)\}_{n=0}^\infty\right)$ is a real-analytic isomorphism between $\bH_+$ and
$\cS_D\times\R\times\el2_{3/4}$, where
$$
 \l_n=\l_n^0+\mu_n\,,\ \ n\ge 0\,,\quad {\rm and}\quad
\cS_D=\biggl\{\{h_n\}_{n=0}^\infty\in\cH:
\lambda_1^0\!+\!h_0\!<\!\lambda_3^0\!+\!h_1\!<\!\lambda_5^0\!+\!h_2\!<\!\dots \biggr\}.
$$
In other words, after this modification we have the spectral mapping with similar properties
as in the case of the real line \cite{CKK} and the new parameters
$\{r_{2n+1}\}_{n=0}^{\infty}\!\in\!\ell_{3/4}^2$ give the correct parameterization of the
isospectral manifolds.

The inverse problem for the family of operators $\{T_b\}_{b\in\R}$ is similar to the case of
$T_D$ but we need to control the dependence of our spectral data from the boundary constant
$b\in\R$\,.  Here Theorem \ref{*KasSD} plays a crucial role. It is similar to the case of the
Sturm-Liouville operators $-y''+\!q(x)y$ on $[0,1]$ with the mixed boundary conditions
$y(0)\!=\!y'(1)\!=\!0$\,, see \cite{K}.

The ingredients of the proof of main theorem are:\\
i) Uniqueness Theorem. We adopt the proof from \cite{PT} and \cite{CKK1}.
This proof requires only some estimates of the fundamental solutions.\\
ii) Analysis of the Fr\'echet derivative of the nonlinear spectral mapping
$\{{\rm potentials}\}\!\mapsto\!\{{\rm spectral\ data}\}$ at the point $q\!=\!0$\,.
We emphasize that this linear operator is complicated (in particular, it is not the Fourier
transform, as it was in \cite{PT}). Here we essentially use the technique of generating
functions (from \cite{CKK}), which are analytic in the unit disc.\\
iii) Asymptotic analysis of the difference between spectral data and its Fr\'echet
derivatives at $q\!=\!0$\,. Here the calculations and asymptotics from \cite{CKK}
play an important role.\\
iv) The proof that the spectral mapping is a surjection, i.e. the fact that each element of an
appropriate Hilbert (or Banach) space can be obtained as spectral data of some potential
$q\!\in\!\bH_+$\,. Here we use the standard Darboux transform of second-order differential
equations.

The present paper continues the series of papers \cite{CKK1}, \cite{CKK} devoted to the
inverse spectral problem for the perturbed harmonic oscillator on the real line. Note that the
set of spectra which correspond to potentials from $\bH_+$ (see Sect. 3 for details) is
similar to the space of spectral data in \cite{CKK}\,. In particular, the range of the linear
operator $f(z)\!\mapsto\!(1\!-\!z)^{-1/2}f(z)$ acting in some Hardy-Sobolev space in the unit
disc plays an important role. As a byproduct of our analysis, we give the simple proof of the
equivalence between two definitions of the space of spectral data which was established in
\cite{CKK} using a more involved technique.

\section{Main results}

We recall some basic results from \cite{CKK}. Consider the operator
$$
T\psi=-\psi''+x^2\psi+q(x)\psi\,,\quad\quad q\in\bH_{even}\!=
\!\biggl\{q\!\in\!L^2(\R):q',xq\!\in\!L^2(\R);\ q(x)\!=\!q(-x),\ x\!\in\!\R \biggr\}\,,
$$
acting in the space $L^2(\R)$. The spectrum $\s(T)$ is an increasing sequence of  simple
eigenvalues given by
$$
\l_n(q)\!=\!\l_n^0\!+\!\mu_n(q)\,,\quad {\rm where}\quad \l_n^0\!=\!\l_n(0)\!=\!2n\!+\!1,\
n\!\ge\!0\,,\quad {\rm  and}\quad  \mu_n(q)\!\to\!0\ {\rm  as} \ n\!\to\!\infty\,.
$$
Define the real weighted $\el2$-space
$$
\el2_r=\biggl\{c=\{c_n\}_{n=0}^\infty: \ c_n\!\in\! \R\,,\
\|c\|_{\el2_r}^2={\textstyle\sum_{n\ge 0}} (1\!+\!n)^{2r}|c_n|^2\!<\!+\infty\biggr\}\,, \ \
r\!\ge\!0\,,
$$
and the Hardy-Sobolev space of functions, which  are analytic in the unit disc
$\D\!=\!\{z:|z|\!<\!1\}$:
$$
H^2_r=H^2_r(\D)= \biggl\{f(z)\!\equiv\!{\textstyle\sum_{n\ge 0}}f_nz^n,\ z\!\in\!\D:\
f_n\!\in\!\R\,,\ \|f\|_{H^2_r}=\|\{f_n\}_{n=0}^\infty\|_{\el2_r}\!<\!+\infty\biggr\}\,, \ \
r\!\ge\!0\,.
$$
Introduce the {\bf space of spectral data}
\begin{equation}
\label{cHDef} \cH=\biggl\{h=\{h_n\}_{n=0}^{\infty}: \ \sum_{n\ge
0}h_nz^n\equiv\frac{f(z)}{\sqrt{1\!-\!z}}\,, \ f\!\in\!H^2_{\frac{3}{4}}\,\biggr\}\,,\quad
\|h\|_{\cH}=\|f\|_{H^2_{{3}/{4}}}\,.
\end{equation}
\begin{theorem}[\cite{CKK}]  \label{*HevenChar}
The mapping $q\mapsto\{\mu_n(q)\}_{n=0}^\infty$ is a real-analytic
isomorphism\begin{footnote}{By definition, the mapping of Hilbert spaces  $F:H_1\to H_2$ is a
local real-analytic isomorphism iff for any $y\!\in\!H_1$ it has an analytic continuation
$\widetilde{F}$ into some complex neighborhood $y\!\in\!U\!\subset\!{H_1}_\C$ of $y$ such that
$\widetilde{F}$ is a bijection between $U$ and some complex neighborhood
$F(y)\!\in\!\widetilde{F}(U)\!\subset\!{H_2}_\C$ of $F(y)$  and both
$\widetilde{F}$\,,\,$\widetilde{F}^{-1}$ are analytic. The local isomorphism $F$ is a (global)
isomorphism iff it is a bijection.}\end{footnote} between the space of even potentials
$\bH_{even}$ and the following open convex subset of $\cH$\,:
$$
\cS=\biggl\{\{h_n\}_{n=0}^\infty\in\cH:
\l_0^0\!+\!h_0\!<\!\l_1^0\!+\!h_1\!<\!\l_2^0\!+\!h_2\!<\!\dots \biggr\}\,.
$$
\remark {\rm The inequalities in the definition of $\cS$ correspond to the condition of
monotonicity of eigenvalues of $T$: $\l_0(q)\!<\!\l_1(q)\!<\!\l_2(q)\!<\!\dots$}
\end{theorem}

Recall  the following more precise representation from \cite{CKK}:
\begin{equation}
\label{IntCH0Asympt}
\l_n(q)=\l_n^0+\mu_n(q),\quad \mu_n(q)=\frac{\int_\R q(t)dt}{\pi
\sqrt{\l_n^0}} + \widetilde{\mu}_n(q),\quad \{\widetilde{\mu}_n(q)\}_{n=0}^\infty\in \cH_0\,,
\end{equation}
where the subspace $\cH_0\!\subset\!\cH$ of codimension $1$ is given by
\begin{equation}
\label{cH0Def} \cH_0= \biggl\{h\!\in\!\cH:\sqrt{1\!-\!z}\,
{\textstyle\sum_{n\ge0}}h_nz^n\big|_{z=1}\!=\!f(1)\!=\!0\biggr\}\,.
\end{equation}
Remark that Lemma \ref{*cH0Asymp} yields
$\widetilde{\mu}_n(q)\!=\!O(n^{-\frac{3}{4}}\log^{\frac{1}{2}}n)$ as $n\!\to\!\infty$\,.

Next we consider the inverse problem for the operator $T_D$ on $\R_+$\,. For each
$q\!\in\!\bH_+$ we put $q(-x)\!=\!q(x)$\,, $x\!\ge\!0$\,. This gives a natural isomorphism
between $\bH_+$ and $\bH_{even}$\,. We have
$$
\sigma(T)=\sigma(T_D)\cup\sigma(T_N)\,, \quad
\sigma(T_N)=\{\lambda_{2n}(q)\}_{n=0}^{\infty}\,,\quad
\sigma(T_D)=\{\lambda_{2n+1}(q)\}_{n=0}^\infty\,.
$$
Using Theorem \ref{*HevenChar} and Lemma \ref{*cHEvenOdd}, we obtain
the following description of the perturbations of $\sigma(T_D)$ and $\sigma(T_N)$ which
correspond to potentials $q\!\in\!\bH_+$:
$$
\cS_D= \biggl\{\{\mu_{2n+1}(q)\}_{n=0}^\infty\,,\,q\!\in\!\bH_+\biggr\}=
\biggl\{\{h_n\}_{n=0}^\infty\in\cH:
\lambda_1^0\!+\!h_0\!<\!\lambda_3^0\!+\!h_1\!<\!\lambda_5^0\!+\!h_2\!<\!\dots \biggr\}\,,
$$
$$
\cS_N=\biggl\{\{\mu_{2n}(q)\}_{n=0}^\infty\,,\,q\!\in\!\bH_+\biggr\}=
\biggl\{\{h_n\}_{n=0}^\infty\in\cH:
\l_0^0\!+\!h_0\!<\!\l_2^0\!+\!h_1\!<\!\l_4^0\!+\!h_2\!<\!\dots \biggr\}=\cS_D
$$
(recall that $\l_{2n+1}^0\!=\!\l_{2n}^0\!+\!2$\,, $n\!\ge\!0$). It is  important that two
spectra $\sigma(T_D)$ and $\sigma(T_N)$ are not independent.
The following Theorem describes this relationship between $\{\mu_{2n+1}\}_{n=0}^\infty$ and
$\{\mu_{2n}\}_{n=0}^\infty$\,. Let
$$
\t_n=\mu_{2n}-\mu_{2n+1}=\l_{2n}-\l_{2n+1}+2\,,\quad n\!\ge\!0\,.
$$
\begin{theorem}
\label{*H+2SpecChar} (i) The mapping $q\mapsto\biggl(\{\mu_{2n+1}(q)\}_{n=0}^{\infty}\,,
\{\t_n(q)\}_{n=0}^\infty\biggr)$ is a real-analytic isomorphism between $\bH_+$ and the
following open and convex subset of $\cH\times\el2_{3/4}$\,:
$$
\biggl\{\biggl(\{h_n\}_{n=0}^\infty\,,\{\t_n\}_{n=0}^\infty\biggr)\in\cH\times\el2_{\frac{3}{4}}:
\l_0^0\!+\!h_0\!+\!\t_0\!<\!\l_1^0\!+\!h_0\!<\!
\l_2^0\!+\!h_1\!+\t_1\!<\!\l_3^0\!+\!h_1\!<\!\dots\biggr\}\,.
$$
(ii) The mapping $q\mapsto\biggl(\{\mu_{2n+1}(q)\}_{n=0}^{\infty}\,,
\{\t_n(q)\}_{n=0}^\infty\biggr)$ is a real-analytic isomorphism between $\bH_+$ and the
following open and convex subset of $\cH\times\el2_{3/4}$\,:
$$
\biggl\{\biggl(\{h_n\}_{n=0}^\infty\,,\{\t_n\}_{n=0}^\infty\biggr)\in\cH\times\el2_{\frac{3}{4}}:
\l_0^0\!+\!h_0\!<\!\l_1^0\!+\!h_0\!-\!\t_0\!<\!
\l_2^0\!+\!h_1\!<\!\l_3^0\!+\!h_1\!-\!\t_1\!\!<\!\dots\biggr\}\,.
$$
(iii) For each $q\in \bf H_+$ the identity $q(0)=2\sum_{n\ge 0}\tau_n(q)$ holds true. \remark
{\rm As in Theorem \ref{*HevenChar}, these subsets are open and convex. Inequalities in its
definitions correspond to the monotonicity of the sequence $\{\l_m(q)\}_{m=0}^\infty$\,.}
\end{theorem}

We now come to the main results of our paper. Namely, we consider the spectrum of $T_D$ and a
convenient choice of additional spectral data (coordinates of an isospectral set). As a first
step, we introduce the norming constants $s_{2n+1}(q)$. In order to keep the connection with
the problem on $\R$ we use odd indices. We consider the unperturbed equation
$$
-\p''+x^2\p=\l\p\,,\quad \l\!\in\!\C\,,
$$
and its solutions $\p_+^0(x,\l)\!=\!D_{\frac{\l-1}{2}}(\sqrt{2}x)$\,, where $D_\mu(x)$ is a
Weber function (see \cite{B}). The perturbed equation
\begin{equation}
\label{PertEq} -\psi''+x^2\psi+q(x)\psi=\l\psi\,,\quad \l\!\in\!\C\,, \ \ q\!\in\!\bH_+,
\end{equation}
has a solution $\psi_+(x,\l,q)$ such that
$\psi_+(x,\lambda,q)\!=\!\psi_+^0(x,\lambda)(1\!+\!o(1))$\,, $x\!\to\!+\infty$ (see
(\ref{Psi+Asympt1})). The norming constant (see also \cite{MT}, \cite{IT}, \cite{PT}) is
defined by
\begin{equation}
\label{SnDefD} s_{2n+1}(q)=-\log|\psi'_+(0,\lambda_{2n+1}(q),q)|=
\log\left[(-1)^n
\frac{\varphi(\cdot,\lambda_{2n+1}(q),q)}{\psi_+(\cdot,\lambda_{2n+1}(q),q)}\right]\,, \quad
n\!\ge\!0\,,
\end{equation}
where $\varphi$ is the solution of (\ref{PertEq}) such that $\varphi(0,\l,q)\!=\!0$ and
$\varphi'(0,\l,q)\!=\!1$\,.

\pagebreak

The standard identities $\psi_+^2\!=\!\{\dot{\psi}_+\,,\psi_+\}'$,
$\varphi^2\!=\!\{\dot{\varphi}\,,\varphi\}'$ imply\begin{footnote}{Here and below we use the
notations $\dot{\phantom{x}}\!=\!\partial\big/\partial\lambda$\,, $\{f,g\}\!=\!fg'\!-\!f'g$\,,
$\|\cdot\|_+\!=\!\|\cdot\|_{L^2(\R_+)}$\,, $(\cdot,\cdot)_+\!=\!(\cdot,\cdot)_{L^2(\R_+)}$\,.
}\end{footnote}
$$
\|\psi_+(\cdot,\lambda_{2n+1}(q),q)\|^2_{+}=
(-1)^n\dot{w}_D(\lambda_{2n+1}(q),q)e^{-s_{2n+1}(q)}\,,
$$
$$
\|\varphi(\cdot,\lambda_{2n+1}(q),q)\|^2_{+}=
(-1)^n\dot{w}_D(\lambda_{2n+1}(q),q)e^{s_{2n+1}(q)}\,,
$$
where
$$
w_D(\lambda,q)=\{\varphi,\psi_+\}(\lambda,q)= -\psi_+(0,\lambda,q)= -\psi_+^0(0,\lambda)\cdot
\prod_{n\ge 0} \frac{\lambda\!-\!\lambda_{2n+1}(q)}{\lambda\!-\!\lambda_{2n+1}^0}\
$$
and the entire function $\psi_+^0(0,\lambda)\!=\!D_{\frac{\lambda-1}{2}}(0)$ is given by
(\ref{Psi+00lambda}).

If $q\!=\!0$, then the norming constants have the form\begin{footnote}{We use the standard
notation $f\sim g \Leftrightarrow f=g(1+o(g))$\,.}\end{footnote}
$$
s_{2n+1}^0=-\log|(\psi^0_+)'(0,\l_{2n+1}^0)|=
-\log\pi^{-\frac{1}{2}}{2^{n+\frac{3}{2}}}\Gamma({\textstyle n\!+\!\frac{3}{2}}) \sim -n\log
n\,,\quad n\!\to\!\infty\,.
$$
In many cases\begin{footnote}{Sturm-Liouville operators $-\frac{d^2}{dx^2}\!+\!q(x)$ on
$[0,1]$ (\cite{PT},\cite{IT},\cite{IMT},\cite{DT}, \cite{K1}) with all types of "separated"\
boundary conditions;
perturbed harmonic oscillator $-\frac{d^2}{dx^2}\!+\!x^2\!+\!q(x)$ on $\R$  (\cite{MT},
\cite{CKK}).}\end{footnote}, the sequence of eigenvalues and the sequence of norming constants
are "independent coordinates"\ in the space of potentials. But, different from most known
cases,  $\{\mu_{2n+1}\}_{n=0}^{\infty}$ and $\{s_{2n+1}\}_{n=0}^\infty$ are not independent
parameters (establishing a real analytic isomorphism between $\bf H_+$ and some Hilbert
space). To find such an isomorphism, we use some {\bf modification}
$\{r_{2n+1}\}_{n=0}^\infty$ of the norming constants $\{s_{2n+1}\}_{n=0}^\infty$\,. This
modification is not trivial and it uses the discrete Hilbert transform. Namely, we define
$r_{2n+1}(q)$ by
\begin{equation}
\label{RnDefD}
s_{2n+1}(q)=s_{2n+1}^0+\alpha_{2n+1}\cdot\mu_{2n+1}(q)+\frac{q(0)}{4(2n\!+\!1)}+
\frac{1}{2}\sum_{m\ge 0} \frac{\mu_{2m+1}(q)}{2(n\!-\!m)\!+\!1}+r_{2n+1}(q)\,,\quad
n\!\ge\!0\,,
\end{equation}
where
$$
\alpha_{2n+1}=-\frac{(\dot{\psi}_+^0)'}{(\psi_+^0)'}\,(0,\lambda_{2n+1}^0)\sim -\log n\,,\quad
n\!\to\!\infty\,.
$$
\begin{theorem} \label{*H+DChar}
(i) Let $q\!\in\!\bH_+$\,. Then $\{r_{2n+1}(q)\}_{n=0}^\infty\!\in\!\el2_{3/4}$\,, where
$r_{2n+1}$ is given by (\ref{RnDefD})\\
(ii) The mapping $q\mapsto \left(\{\mu_{2n+1}(q)\}_{n=0}^\infty\,,q(0)\,,
\{r_{2n+1}(q)\}_{n=0}^\infty\right)$ is a real-analytic isomorphism between $\bH_+$ and
$\cS_D\times\R\times\el2_{3/4}$\,. \remark {\rm $\left(q(0)\,,
\{r_{2n+1}(q)\}_{n=0}^\infty\right)$ are "independent coordinates"\ in the isospectral sets.
}
\end{theorem}

We consider the inverse problem for the family of operators $T_by\!=\!-y''\!+\!x^2y\!+\!q(x)y$
with the boundary conditions $y'(0)\!=\!b y(0)$\,, $b\!\in\!\R$\,. Denote its spectrum by
$$
\sigma(T_b)=\{\l_{2n}(q,b)\}_{n=0}^\infty= \{\l_{2n}^0\!+\!\mu_{2n}(q,b)\}_{n=0}^\infty
$$
(note that $\lambda_{2n}(q,0)\!=\!\lambda_{2n}(q)$). Then (see (\ref{LSAsymptN}))
\begin{equation}
\label{IntCH0AsymptN} \lambda_{2n}(q,b)=\l_{2n}^0+\frac{2\int_{\R_+}q(t)dt+2b}
{\pi\sqrt{\l_{2n}^0}}+ \widetilde{\mu}_{2n}(q,b)\,,\quad
\{\widetilde{\mu}_{2n}(q,b)\}_{0}^\infty\!\in\!\cH_0,
\end{equation}
for $(q,b)\!\in\!\bH_+\!\times\!\R$ and, in particular,
$\widetilde{\mu}_{2n}(q,b)\!=\!O(n^{-\frac{3}{4}}\log^{\frac{1}{2}}n)$\,, $n\!\to\!\infty$\,.

\pagebreak

By analogy with (\ref{SnDefD}), (\ref{RnDefD}) we introduce the norming constants
$s_{2n}(q,b)$ and the new parameters $r_{2n}(q,b)$ by
\begin{equation}
\label{SnDefN} s_{2n}(q,b)=-\log|\psi_+(0,\l_{2n}(q),q)|= \log\left[(-1)^n
\frac{(\vartheta\!+\!b\varphi)(\cdot,\lambda_{2n}(q),q)}
{\psi_+(\cdot,\lambda_{2n}(q),q)}\right]
\end{equation}
\begin{equation}
\label{RnDefN} =s_{2n}^0+\alpha_{2n}\cdot\mu_{2n}(q,b) - \frac{q(0)-2b^2}{4(2n\!-\!1)} +
\frac{1}{2}\sum_{m\ge 0} \frac{\mu_{2m}(q,b)}{2(n\!-\!m)\!-\!1}+r_{2n}(q,b)\,,\quad
n\!\ge\!0\,,
\end{equation}
where $\vartheta$ is the solution of (\ref{PertEq}) such that $\vartheta(0,\l,q)\!=\!1$\,,
$\vartheta'(0,\l,q)\!=\!0$ and
$$
s_{2n}^0=-\log|\psi^0_+(0,\l_{2n}^0)| \sim -n\log n\,,\quad
\alpha_{2n}=-\frac{\dot{\psi}_+^0}{\psi_+^0}\,(0,\l_{2n}^0)\sim-\log n\,, \quad
n\!\to\!\infty\,.
$$
\begin{theorem} \label{*H+NChar}
(i) Let $(q,b)\!\in\!\bH_+\!\times\!\R$. Then $\{r_{2n}(q,b)\}_{n=0}^\infty\!\in\!\el2_{3/4}$,
where $r_{2n}$ is given by (\ref{RnDefN}).\\
(ii) The mapping $(q,b)\mapsto \left(\{\mu_{2n}(q,b)\}_{n=0}^\infty\,,q(0)\!-\!2b^2,
\{r_{2n}(q,b)\}_{n=0}^\infty\right)$ is a real-analytic isomorphism between $\bH_+\!\times\R$
and $\cS_N\times\R\times\el2_{3/4}$\,.
\end{theorem}

Finally, we give the explicit expression of the "boundary"\ constant $b$ in terms of spectral
data. Introduce the Wronskian
$$
w_N(\l,q,b)=\p'_+(0,\l,q)-b\p_+(0,\l,q)\,,\quad \l\!\in\!\C\,.
$$
Note that the entire function $w_N(\lambda,q,b)$ can be reconstructed by its zeros
$\{\lambda_{2n}(q,b)\}_{n=0}^\infty$ as
$$
w_N(\l,q,b)= (\psi^0_+)'(0,\l)\cdot \prod_{n\ge 0}
\frac{\lambda\!-\!\lambda_{2n}(q,b)}{\l\!-\!\l_{2n}^0}\,,\quad \l\!\in\!\C\,,
$$
where the entire function $(\psi_+^0)'(0,\lambda)\!=\!\sqrt{2}D'_{\frac{\lambda-1}{2}}(0)$ is
given by (\ref{Psi+00lambda}).
\begin{theorem}
\label{*KasSD} For each $(b,q)\!\in\! \R\!\times\!\bH_+$ the following identity is fulfilled:
\begin{equation}
\label{KisEqual} -b=\sum_{n\ge 0} \biggl(\frac{(-1)^ne^{-s_{2n}(q,b)}}
{\dot{w_N}(\l_{2n}(q,b),q,b)}\,-2\pi^{-1/2}E_n\biggr)\,, \quad
E_n\!=\frac{(2n)!}{2^{2n}(n!)^2}\,.
\end{equation}
\remark {\rm In the proof we use ideas from \cite{K}.}
\end{theorem}

The plan of the paper is as follows. In Section 3 we recall some results of \cite{CKK},
describe basic properties of $\cH$ and prove Theorem \ref{*H+2SpecChar}. Section 4 is devoted
to the Dirichlet boundary condition at $x\!=\!0$ (here we prove Theorem \ref{*H+DChar}). In
Section 5 we solve the inverse problem for the family of the operators $\{T_b\}_{b\in\R}$
(Theorems \ref{*H+NChar}, \ref{*KasSD}). All needed properties of fundamental solutions,
gradients of spectral data and asymptotics are collected in Appendix.

\section{Preliminaries and proof of Theorem \ref{*H+2SpecChar}}
\setcounter{equation}{0}

We recall some crucial points of \cite{CKK}. Let $\p_n^0$ be the normalized eigenfunctions of
the unperturbed harmonic oscillator on $\R$. It is well-known that
$$
\psi_n^0(x)=(n!\sqrt{\pi})^{-\frac{1}{2}}D_n(\sqrt{2}x)=
({2^n}n!\sqrt{\pi})^{-\frac{1}{2}}H_n(x)e^{-\frac{x^2}{2}}\,,\quad n\!\ge\!0\,,
$$
where $H_n(x)$ are the Hermite polynomials. For each $n\!\ge\!0$ we consider the second
solution
$$
\chi_n^0(x)=\biggl(\frac{n!\sqrt{\pi}}{2}\biggr)^{\!1/2}
\cases{(-1)^\frac{n}{2}\Im{D_{-n-1}}(i\sqrt{2}x), & $n$ is even, \cr
(-1)^\frac{n-1}{2}\Re{D_{-n-1}}(i\sqrt{2}x), & $n$ is odd,}
$$
of the equation $-y''\!+\!x^2y\!=\!\lambda_n^0 y$\,, which is uniquely defined by the
conditions
$$
\{\chi_n^0\,,\psi_n^0\}\!=\!1\,,\qquad (\psi_n^0\chi_n^0)(-x)\!=\!-(\psi_n^0\chi_n^0)(x)\,,\ \
\ x\in \R\,.
$$
Note that $(\psi_n^0\chi_n^0)(x)\sim (-1)^{n+1}x$ as $x\!\to\!0$\,,  and
$(\psi_n^0\chi_n^0)(x)\sim -x^{-1}$ as $x\!\to\!\infty$\,.

Let $s_{2m}(q)\!=\!s_{2m}(q,0)$\,, $m\!\ge\!0$\,. Define the modified norming constants
$\widetilde s_{n}(q)$ by
\begin{equation}
\label{TildeSnDef} s_{n}(q)=s_{n}^0+\alpha_{n}\mu_{n}(q)+\widetilde s_{n}(q)\,,\quad n\!\ge\!0
\end{equation}
(recall that $s_n^0\!=\!s_n(0)$\,,
$\alpha_{2m+1}\!=\!-{(\dot{\psi}_+^0)'}\big/{(\psi_+^0)'}\,(0,\lambda_{2m+1}^0)$ and
$\alpha_{2m}=-{\dot{\psi}_+^0}\big/{\psi_+^0}\,(0,\l_{2m}^0)$\,, $m\!\ge\!0$).

\begin{theorem}
\label{*Asympt} For each potential $q\!\in\!\bH_+$ the following asymptotics are
fulfilled\begin{footnote}{Here and below $a_n\!=\!b_n\!+\!\el2_r(n)$ means that
$\{a_n\!-\!b_n\}_{n=0}^\infty\!\in\!\el2_r$\,. We say that $a_n(q)\!=\!b_n(q)\!+\!\el2_r(n)$
holds true uniformly on some set iff norms $\|\{a_n(q)\!-\!b_n(q)\}_{n=0}^\infty\|_{\el2_r}$
are uniformly bounded on this set.}\end{footnote}:
\begin{equation}
\label{LnSnAsympt} \mu_n(q)=2{\nh q_n^+}+\el2_{\frac{3}{4}+\delta}(n)\,,\qquad
\widetilde{s}_n(q)= {\nc q_n^+}+\el2_{\frac{3}{4}+\delta}(n)\,,
\end{equation}
for some absolute constant $\delta\!>\!0$,
uniformly on bounded subsets of $\bH_+$, where
$$
\nh q_n^+=(q,(\psi_n^0)^2)_{+}\,,\qquad \nc q_n^+ =(q,\psi_n^0\chi_n^0)_{+}\,.
$$
\remark {\rm i) This result does not cover the general boundary condition $y'(0)\!=\!b
y(0)$\,,
$b\!\ne\!0$\,.\\
ii) Using Proposition \ref{*NcNhasFG} and Corollary \ref{*H0EquivDef}, it is easy to see that
$\{\nh q_n^+\}_{n=0}^{\infty}\!\in\!\cH$\,, $\{\nc q_n^+\}_{n=0}^{\infty}\!\in\!\cH_0$ for
each $q\!\in\!\bH_+$\,. Then, Lemma \ref{*cH0Asymp} and Proposition \ref{*H0Emb} (i) yield
$$
\textstyle \nh q_n^+=\pi^{-1}\int_{\R_+}q(t)dt\cdot (\lambda_n^0)^{-\frac{1}{2}}
+O(n^{-\frac{3}{4}}\log^{\frac{1}{2}}n)\,,\qquad \nc
q_n^+=O(n^{-\frac{3}{4}}\log^{\frac{1}{2}}n)\,,\quad n\!\to\!\infty\,.
$$}
\end{theorem}
\begin{proof} Let $\mu_{n}\!=\!\mu_{n}(q)$. We suppose that $n$ is odd,
the other case is similar. Repeating arguments of \cite{CKK1} Lemma 4.2 and using Rouch\'e's
Theorem, we obtain $\mu_{n}(q)\!=\!O(n^{-1/2})$\,. Note that the value $\mu_{n}$ is the
solution of the equation $\psi_+(0,\lambda_{n}^0\!+\!\mu_{n},q)\!=\!0$\,. Due to estimates
from Corollary \ref{*ADotEstim} and asymptotics (\ref{KappaValues}), we have
$$
0=\frac{\psi_+(0,\lambda_{n}^0\!+\!\mu_{n},q)}{\dot{\kappa}_{n}} =
\frac{\psi_+^{(1)}(0,\lambda_{n}^0,q)+\dot{\kappa}_{n}\cdot\mu_{n}} {\dot{\kappa}_{n}} +
O(n^{-1}\log^2n).
$$
Hence, Lemma \ref{*LA511} (ii) gives $\mu_{n} = 2\nh q_{n}^+ + O(n^{-1}\log^2n)$. Let
$\mu_{n}^{(1)} = 2\nh q_{n}^+$\,. Using the similar arguments, we obtain
$$
0=\!\frac{1}{\dot{\kappa}_{n}}\,\biggl(\psi_+^{(1)} \!+ \psi_+^{(2)} \!+
\dot{\psi}_+^0\cdot\mu_{n} + \dot{\psi}_+^{(1)}\cdot \mu_{n}^{(1)} +
\frac{\ddot{\psi}_+^0}{2}\,(\mu_{n}^{(1)})^2\biggr)(0,\lambda_{n}^0,q) +
O(n^{-\frac{3}{2}}\log^3n).
$$
Together with Lemma \ref{*LA511} (ii), it follows
$$
\frac{\mu_{n}}{2}= \nh q_{n}^+ - \nc q_{n}^+\nh q_{n}^+ +
\biggl(\frac{\dot{\kappa}'_{n}}{\kappa'_{n}}\,\nh q_{n}^+ \!+\! \frac{1}{2}\,\nc
q_{n}^+\biggr)\cdot 2\nh q_{n}^+ -\frac{\ddot{\kappa}_{n}}{\kappa_{n}}\,(\nh q_{n}^+)^2 +
\el2_{\frac{3}{4}+\delta}(n)
$$
$$
= \nh q_{n}^+ + \biggl(\frac{2\dot{\kappa}'_{n}}{\kappa'_{n}} -
\frac{\ddot{\kappa}_{n}}{\kappa_{n}}\biggr)\cdot (\nh q_{n}^+)^2 +\el2_{\frac{3}{4}+\delta}(n)
= \nh q_{n}^+ + \el2_{\frac{3}{4}+\delta}(n),
$$
where we have used (\ref{KappaSpecAsymp}).

Furthermore, we have
$$
e^{s_{n}^0-s_{n}(q)}= \frac{\psi'_+(0,\l_{n}(q),q)}{\kappa'_{n}}= 1- \nc q_{n}^+ +
\frac{\dot{\kappa}'_{n}}{\kappa'_{n}}\,\cdot \mu_{n} + \frac{1}{2}\,(\nc q_{n}^+)^2 \!-\!
\frac{\pi^2}{8}\,(\nh q_{n}^+)^2
$$
$$
+ \biggl(- \frac{\dot{\kappa}'_{n}}{\kappa'_{n}}\,\nc q_{n}^+ \!+\! \frac{\pi^2}{8}\,\nh
q_{n}^+\biggr)\cdot \mu_{n}^{(1)} +
\frac{\ddot{\kappa}'_{n}}{2\kappa'_{n}}\,(\mu_{n}^{(1)})^2+ \el2_{\frac{3}{4}+\delta}(n)\,.
$$
Hence,
$$
s_{n}^0-s_{n}(q) = -\nc q_{n}^+ + \frac{\dot{\kappa}'_{n}}{\kappa'_{n}}\,\cdot \mu_{n}
+\biggl(\frac{\ddot{\kappa}'_{n}}{\kappa'_{n}}-
\frac{(\dot{\kappa}'_{n})^2}{(\kappa'_{n})^2}+\frac{\pi^2}{16}\biggr)\cdot
\frac{(\mu_{n}^{(1)})^2}{2}+\el2_{\frac{3}{4}+\delta}(n)
$$
$$
=-\nc q_{n}^+ + \frac{\dot{\kappa}'_{n}}{\kappa'_{n}}\,\cdot
\mu_{n}+\el2_{\frac{3}{4}+\delta}(n),
$$
where we have used (\ref{KappaSpecAsymp}). This yields the second part of (\ref{LnSnAsympt})
in view of (\ref{TildeSnDef}).
\end{proof}

Below we need some special representation of the sequences  $\{\nh q_n^+\}_{n=0}^{\infty}$ and
$\{\nc q_n^+\}_{n=0}^\infty$\,. Define the functions
$$
\widetilde{\psi}_n^0(x)= 2^{1/4}\psi_n^0(\sqrt{2}x)\,,\quad n\!\ge\!0\,.
$$
Note that the system of functions $\{\widetilde{\psi}_{2m}^0\}_{m=0}^\infty$ is the
orthogonal basis of $\bH_{even}$\,, i.e. the orthogonal basis of the space $\bH_+$\,, while
the system $\{\widetilde{\psi}_{2m+1}^0\}_{m=0}^\infty$ is the orthogonal basis of its
subspace $\bH_+^0=\{q\!\in\!\bH_+:q(0)\!=\!0\}\subsetneq\bH_+$\,.

Following \cite{CKK}, for each $q\in \bH_+$ we define two (analytic in the unit disc $\D$)
functions
\begin{equation}\label{FqGqDef}
\begin{array}{c}\displaystyle
(F^+q)(z)\equiv\frac{1}{(2\pi)^{1/4}}\sum_{k\ge 0}
\sqrt{E_k}(q,\widetilde{\psi}_{2k}^0)_{+}\cdot z^k\,,\quad z\!\in\!\D\,,\cr \displaystyle
(G^+q)(z)\equiv-\frac{(2\pi)^{1/4}}{2}\sum_{k\ge 0}
\frac{(q,\widetilde{\psi}_{2k+1}^0)_{+}}{\sqrt{(2k\!+\!1)E_k}}\,z^k\,,\quad z\!\in\!\D\,,
\end{array}
\end{equation}
where
$$
E_k=\frac{(2k)!}{2^{2k}(k!)^2}\sim \pi^{-\frac{1}{2}}k^{-\frac{1}{2}}\quad {\rm as}\quad
k\!\to\!\infty\,.
$$

\pagebreak

Note that $F^+q\,,G^+q\!\in\!H^2_{3/4}$ for each $q\!\in\!\bH_+$\,. Moreover, the mapping
$q\mapsto F^+q$ is a linear isomorphism between $\bH_+$ and $H^2_{3/4}$\,, while the mapping
$q\mapsto G^+q$ is a linear isomorphism between $\bH_+^0$ and $H^2_{3/4}$\,. Recall that $\nh
q_n^+=\nolinebreak (q,(\psi_n^0)^2)_{+}$ and $\nc q_n^+ =(q,\psi_n^0\chi_n^0)_{+}$\,,
$n\!\ge\!0$\,.

\begin{proposition}
\label{*NcNhasFG} For each $q\!\in\!\bH_+$ the following identities\begin{footnote}{We write
$f(z)\equiv g(z)$ iff the identity $f(z)=g(z)$ holds true for all
$z\!\in\!\D$\,.}\end{footnote} are fulfilled:
\begin{equation}
\label{NcNhAsFqGq} \sum_{n\ge 0} \nh q_n^+z^n \equiv\frac{(F^+q)(z)}{\sqrt{1\!-\!z}}\,,\qquad
\sum_{n\ge 0} \nc q_n^+z^n \equiv
P_+\biggl[\frac{(G^+q)(\zeta)}{\sqrt{1\!-\!\overline{\zeta}}}\biggr]\,,\qquad z\in\D\,,
\end{equation}
\begin{equation}
(F^+q)(1)\!=\!(2\pi)^{-\frac{1}{2}}\int_{\R_+}q(t)dt,\qquad
(F^+q)(-1)\!=\!2^{-\frac{3}{2}}q(0)\,.
\end{equation}
\remark {\rm Here and below we put $(P_+f)(z)\equiv\frac{1}{2\pi i}
\int_{|\zeta|=1}\frac{f(\zeta)d\zeta}{\zeta-z}$ for any $f\!\in\!L^1(\T)$ and $z\!\in\!\D$\,.
In particular, the identity $(P_+\sum_{n=-k}^k c_n{\zeta}^n)(z)\!\equiv\!\sum_{n=0}^kc_nz^n$
holds true for any $c_n\!\in\!\C$\,.}
\end{proposition}
\begin{proof}
Identities (\ref{NcNhAsFqGq}) were proved in \cite{CKK} (Propositions 1.2 and 2.9). Also, in
\cite{CKK} it was shown that
$$
1=\sum_{k\ge 0}\widetilde\psi_{2k}^0(x)\int_\R\!\widetilde{\psi}_{2k}^0(t)dt =
(2\pi)^{\frac{1}{4}}\sum_{k\ge 0}\sqrt{E_k}\,\widetilde{\psi}_{2k}^0(x)\,, \quad x\in \R\,,
$$
in the sense of distributions, which gives
$(F^+q)(1)\!=\!(2\pi)^{-\frac{1}{2}}\int_{\R_+}q(t)dt$\,. Furthermore,
$$
\delta(x)=\sum_{k\ge 0}\widetilde{\psi}_{2k}^0(x)\cdot\widetilde\psi_{2k}^0(0) = \sum_{k\ge
0}\widetilde{\psi}_{2k}^0(x)\cdot \frac{2^{1/4}H_{2k}(0)}{(\sqrt{\pi}\,2^{2k}(2k)!)^{1/2}}=
\biggl(\frac{2}{\pi}\biggr)^{\!1/4}\sum_{k\ge 0}(-1)^k\sqrt{E_k}\,\widetilde{\psi}_{2k}^0(x)
$$
in the sense of distributions. Together with (\ref{FqGqDef}), this implies
$(F^+q)(-1)\!=\!2^{-\frac{1}{2}}\cdot\frac{q(0)}{2}$\,.
\end{proof}

Emphasize that definition (\ref{cHDef}) of the space $\cH$ is directly motivated by Eq.
(\ref{NcNhAsFqGq}). The following Proposition gives the basic properties of $\cH$\,, $\cH_0$
(see \cite{CKK} Lemmas 2.10, 2.11).

\begin{proposition}
\label{*H0Emb} (i) For each $\{h_n\}_{n=0}^\infty\!\in\!\cH$ there is a unique decomposition
$h_n\!=\!\frac{v}{\sqrt{\l_n^0}}+h_n^{(0)}$, where
$v\!=\!({\pi}/{2})^{-\frac{1}{2}}\,\sqrt{1\!-\!z}\,\sum_{n=0}^{+\infty} h_nz^n\big|_{z=1}$ and
$\{h_n^{(0)}\}_{n=0}^\infty\!\in\!\cH_0$\,. The mapping $h\mapsto(v,h^{(0)})$ is a linear
isomorphism between $\cH$ and $\R\times\cH_0$\,. If $h\!=\!\{\nh q_n^+\}_{n=0}^\infty$\,,
then $v\!=\!\pi^{-1}\!\int_{\R_+} q(t)dt$\,.\\
(ii) The set of finite sequences $\left\{(h_0\,,\dots\,,h_k\,,0\,,0\,,\dots),\ k\!\ge\!0\,,\
h_j\!\in\!\R\right\}$
is dense in $\cH_0$\,. \\
(iii) The embeddings $\el2_{3/4}\!\subset\!\cH_0\!\subset\!\el2_{1/4}$ are fulfilled. \remark
{\rm Since $\cH_0\!\subset\!\el2_{1/4}$\,, the sequence of leading terms
$\{v\big/\sqrt{\lambda_n^0}\,\}_{n=0}^{\infty}$ doesn't belong to $\cH_0$\,.}
\end{proposition}

The next Lemma gives the $O$-type estimate for elements of $\cH_0$\,.

\begin{lemma} \label{*cH0Asymp} Let $h\!=\!\{h_n\}_{n=0}^\infty\!\in\!\cH_0$\,. Then
$h_n=O(n^{-\frac{3}{4}}\log^{\frac{1}{2}}n)$ as $n\!\to\!\infty$\,.
\end{lemma}
\begin{proof}
The proof is similar to the proof of Lemma 2.1 in \cite{Ch}.
Definition (\ref{cH0Def}) of $\cH_0$ yields
$$
\sum_{n\ge 0}h_nz^n\equiv\frac{\sum_{k\ge 0}f_kz^k}{\sqrt{1\!-\!z}}\,,\quad
\{f_k\}_{k=0}^{\infty}\!\in\!\el2_{3/4}\,,\ \ \sum_{k\ge 0}f_k\!=\!f(1)\!=\!0\,.
$$
Recall that $(1\!-\!z)^{-\frac{1}{2}}\!\equiv\!\sum_{m\ge 0}E_mz^m$\,. Hence,
$$
h_n=\sum_{k=0}^nE_{n-k}f_k=\sum_{k=1}^n(E_{n-k}\!-\!E_n)f_k-E_n\sum_{k=n+1}^{\infty}f_k\,.
$$
It is easy to see that $E_n\!=\!O(n^{-\frac{1}{2}})$ and
$E_{n-k}\!-\!E_n\!=\!O(kn^{-1}(n\!-\!k\!+\!1)^{-\frac{1}{2}})$\,. Therefore,
$$
\biggl|E_n\sum_{k=n+1}^{\infty}f_k\biggr| \le
E_n\biggl(\sum_{k=n+1}^{+\infty}k^{-\frac{3}{2}\,}\biggr)^{\!\!1/2}
\biggl(\sum_{k=n+1}^{+\infty}k^{\frac{3}{2}}|f_k|^2\biggr)^{\!\!1/2}=O(n^{-\frac{3}{4}})\,,
$$
$$
\biggl|\sum_{k=1}^n(E_{n-k}\!-\!E_n)f_k\biggr|\le \biggl(\sum_{k=1}^n
O\biggl(\frac{k^{\frac{1}{2}}}{n^2(n\!-\!k\!+\!1)}\biggr)\,\biggr)^{\!\!1/2}
\biggl(\sum_{k=1}^nk^{\frac{3}{2}}|f_k|^2\biggr)^{\!\!1/2} =
O(n^{-\frac{3}{4}}\log^{\frac{1}{2}}n)\,,
$$
where the inequality $\sum_{k=1}^{+\infty}k^{\frac{3}{2}}|f_k|^2\!<\!+\infty$ has been used.
\end{proof}

 In order to prove Theorem \ref{*H+2SpecChar} we need
\begin{lemma}
\label{*cHEvenOdd} (i) Let $h\!=\!\{h_n\}_{n=0}^\infty\!\in\!\cH$\,. Then
$h_N\!=\!\{h_{2n}\}_{n=0}^\infty\,,\  h_D\!=\!\{h_{2n+1}\}_{n=0}^\infty \!\in\! \cH$ and $\dlt
h\!=\!\{h_{2n}\!-\!h_{2n+1}\}_{n=0}^\infty\!\in\!\el2_{3/4}$\,. \\
(ii)  Both mappings
$h\mapsto(h_N\,,\dlt h)$, $h\mapsto(h_D\,,\dlt h)$ are isomorphisms between $\cH$ and
$\cH\times\el2_{3/4}$\,.
\end{lemma}
\begin{proof}
(i) For $z\!\in\!\D$ denote $h(z)\!\equiv\sum_{n\ge 0}h_nz^n$\,, $h_N(z)\!\equiv\!\sum_{n\ge
0}h_{2n}z^n$ and so on. By definition of $\cH$, we have
$$
h_N(z^2)+zh_D(z^2)\equiv h(z)\equiv\frac{f(z)}{\sqrt{1\!-\!z}}\,,\quad f\!\in\!H^2_{3/4}\,.
$$
Then,
$$
h_N(z^2)\equiv\frac{f_N(z^2)}{\sqrt{1\!-\!z^2}}\,,\quad {\rm where} \quad
f_N(z^2)\equiv\frac{f(z)\sqrt{1\!+\!z}+f(-z)\sqrt{1\!-\!z}}{2}\,,
$$
\begin{equation}
\label{FnFdDIdent} h_D(z^2)\equiv\frac{f_D(z^2)}{\sqrt{1\!-\!z^2}}\,,\quad {\rm where} \quad
f_D(z^2)\equiv\frac{f(z)\sqrt{1\!+\!z}-f(-z)\sqrt{1\!-\!z}}{2z}\,,
\end{equation}
$$
(\dlt h)(z^2)\equiv h_N(z^2)-h_D(z^2)\equiv
\frac{-f(z)\sqrt{1\!-\!z}+f(-z)\sqrt{1\!+\!z}}{2z}\,.
$$
Due to $\sqrt{1\!\pm\!z}\in H^2_{3/4}$\,, we obtain $f_N\,,f_D\,,\dlt h\!\in\!H^2_{3/4}$\,,
and hence $h_N\,,h_D\!\in\!\cH$\,.\\
(ii) By definition of $\cH$, the mappings $h\mapsto f$\,, $h_N\mapsto f_N$ and $h_D\mapsto
f_D$ are isomorphisms between $\cH$ and $H^2_{3/4}$\,. We show that both mappings
$f\mapsto(f_N\,,\dlt h)$\,, $f\mapsto(f_D\,,\dlt h)$ are isomorphisms between $H^2_{3/4}$ and
$H^2_{3/4}\times H^2_{3/4}$\,. Indeed, due to (\ref{FnFdDIdent}), the direct mappings are
bounded. On the other hand, identities
$$
f(z)\equiv f_D(z^2)\sqrt{1\!+\!z}+\dlt h(z^2)\sqrt{1\!-\!z}\equiv f_N(z^2)\sqrt{1\!+\!z}-\dlt
h(z^2)z\sqrt{1\!-\!z}
$$
follow that the inverse mappings are bounded too.
\end{proof}

\begin{proof}[{\bf Proof of Theorem \ref{*H+2SpecChar}.}] Theorem \ref{*HevenChar}
and the second part of Lemma \ref{*cHEvenOdd} give that both mappings
$q\mapsto\left(\{\mu_{2n+1}(q)\}_{n=0}^{\infty}\,, \{\t_n(q)\}_{n=0}^\infty\right)$ and
$q\mapsto\left(\{\mu_{2n}(q)\}_{n=0}^{\infty}\,, \{\t_n(q)\}_{n=0}^\infty\right)$ are
real-analytic isomorphisms. We show that $q(0)=2\sum_{n\ge 0}\t_n(q)$. Proposition
\ref{*NcNhasFG} yields
\begin{equation}
\label{Fq(-1)} \sum_{n\ge 0}({\nh q_{2n}^+}\!-\!{\nh q_{2n+1}^+})=
\frac{(F^+q)(z)}{\sqrt{1\!-\!z}}\Big|_{z=-1}=\frac{q(0)}{4}\,.
\end{equation}
Hence, it is sufficient to check that
\begin{equation}
\label{TraceNonLin} \sum_{n\ge 0}(\mu_{2n}(q)-2\nh q_{2n}^+)=0\,, \qquad \sum_{n\ge 0}
(\mu_{2n+1}(q)-2\nh q_{2n+1}^+)=0
\end{equation}
(note that these series converge due to Theorem \ref{*Asympt}). Below we prove the first
identity, the proof of the second is similar.

Put $q(-x)\!=\!q(x)$, $x\!\ge\!0$\,, and consider the harmonic oscillator on $\R$ perturbed by
the potential $q(x)$\,. The perturbation theory (see e.g. \cite{Ka})
yields\begin{footnote}{Here and below $\partial\xi(q)\big/\partial q\!=\!\zeta(q)$ means that
for any $v\!\in\!L^2$ the equation $(d_q\xi)(v)\!=\!(v,\overline{\zeta})_{L^2}$ holds true.
}\end{footnote}
$$
\frac{\partial\l_{2n}(q)}{\partial q(x)}=\psi_{2n}^2(x,q)\,,\qquad
\frac{\partial\psi_{2n}(x,q)}{\partial q(y)}= \psi_{2n}(y,q)\sum_{m:m\ne 2n}
\frac{\psi_{m}(y,q)\psi_m(x,q)}{\l_{2n}(q)\!-\!\l_m(q)}\,,
$$
where $\psi_{m}(t,q)$\,, $m\!\ge\!0$\,, are the normalized eigenfunctions of the operator $T$.

Therefore,
$$
\mu_{2n}(q)\!-\!2\nh q_{2n}^+= \int_0^1\!\left(q,\psi_{2n}^2(sq)\right)_{L^2(\R)}ds\,-\,\nh
q_{2n}= \int_0^1\left(q,\psi_{2n}^2(sq)\!-\!(\psi_{2n}^0)^2)\right)_{L^2(\R)}ds
$$
and
$$
\left(q(x),\psi_{2n}^2(x,sq)\!-\!(\psi_{2n}^0)^2(x)\right)_{L^2(\R\,,\,dx)}=
\int_0^s\biggl(q(x)\,,\frac{d}{dt}\,\psi_{2n}^2(x,tq)\biggr)_{\!L^2(\R\,,\,dx)}dt
$$
$$
= \int_0^s \biggl(q(x)\,,\, 2\psi_{2n}(x,tq)\!\cdot\!\biggl(q(y)\,,\psi_{2n}(y,tq)
\sum_{m\ne 2n}
\frac{\psi_{m}(y,tq)\psi_m(x,tq)}{\l_{2n}(tq)\!-\!\l_m(tq)}\biggr)_{\!L^2(\R\,,\,dy)}\,
\biggr)_{\!L^2(\R\,,\,dx)}dt\,.
$$
Since $\left(q,(\psi_{2n}\psi_{2m+1})(\cdot,tq)\right)_{L^2(\R)}\!=\!0$ for any
$n,m\!\ge\!0$\,, we have
$$
\sum_{n\ge 0} (\mu_{2n}(q)\!-\!2\nh q^+_{2n})= 2\int_0^1 ds\int _0^s \sum_{n\ge 0}
\sum_{m:m\ne n} \frac{\left(q,(\psi_{2n}\psi_{2m})(tq)\right)_{L^2(\R)}^2}
{\l_{2n}(tq)\!-\!\l_{2m}(tq)}\,dt\,.
$$
Let
$$
S_k=\sum_{n=0}^k\sum_{m:m\ne n}
\frac{\left(q,(\psi_{2n}\psi_{2m})(tq)\right)_{L^2(\R)}^2}{\l_{2n}(tq)\!-\!\l_{2m}(tq)} =
\sum_{n=0}^k\sum_{m=k+1}^{+\infty}
\frac{\left(q,(\psi_{2n}\psi_{2m})(tq)\right)_{L^2(\R)}^2}{\l_{2n}(tq)\!-\!\l_{2m}(tq)}
$$
Using Lemma \ref{*ALemma} and the estimate
$|\lambda_{2n}(tq)\!-\!\lambda_{2m}(tq)|\!\ge\!c(m\!-\!n)$ for some $c\!>\!0$ and
$m\!>\!n\!\ge\!0$\,, we obtain $S_k\!\to\!0$ as $k\!\to\!\infty$\,, which yields $\sum_{n\ge
0}(\mu_{2n}(q)-2\nh q_{2n}^+)=0$\,.
\end{proof}

\section{Inverse problem for the operator $T_D$}
\setcounter{equation}{0}

We begin with the connection between two functions $F^+q$ and $G^+q$ (given by
(\ref{FqGqDef})) for fixed $q\!\in\!\bH_+$. Recall that the system of functions
$\{\widetilde{\psi}_{2k}^0\}_{k=0}^\infty$ is a basis of $\bH_+$. Therefore, it is possible to
rewrite$G^+q$ in terms of $F^+q$\,. For $\zeta\!\in\!\T$\,, $\zeta\!\ne\!-1$\,, we put
$\sqrt{\zeta}\!=\!{\sqrt{e^{i\phi}}}\!=\!e^{\frac{i\phi}{2}}$\,, where $\zeta=e^{i\phi}$,
$\phi\!\in\!(-\pi,\pi)$\,. We have the following identity in $L^2(\T)$:
\begin{equation}
\label{SqrtDef} \frac{1}{\sqrt{\zeta}}=
\frac{2}{\pi}\sum_{l=-\infty}^{+\infty}\frac{(-1)^l}{2l\!+\!1}\,\zeta^l\,, \quad
\zeta\!\in\!\T\,.
\end{equation}
\begin{lemma}
\label{*GasF} For each $q\!\in\!\bH_+$ the following identity
is fulfilled:
\begin{equation}
\label{GasF} (G^+q)(z)\equiv
-\frac{\pi}{2}P_+\biggl(\frac{(F^+q)(\zeta)}{\sqrt{\zeta}}\biggr)\,,\quad z\!\in\!\D\,.
\end{equation}
\end{lemma}
\begin{proof}
We determine the coefficients of the function $\p_{2m+1}^0$ with respect to the basis
$\{\p_{2k}^0\}_{k=0}^\infty$. The standard identity
$\{\psi_{2k}^0\,,\psi_{2m+1}^0\}'\!=\!
(\lambda_{2k}^0\!-\!\lambda_{2m+1}^0)\psi_{2k}^0\psi_{2m+1}^0$\,, yields
$$
({\p}_{2m+1}^0\,,{\psi}_{2k}^0)_{+}= \int_{\R_+}\p_{2m+1}^0(x)\psi_{2k}^0(x)dx=
\frac{\{\p_{2k}^0\,,\p_{2m+1}^0\}(0)}{\l_{2m+1}^0\!-\!\l_{2k}^0}=
\frac{\p_{2k}^0(0)(\p_{2m+1}^0)'(0)}{2(2(m\!-\!k)\!+\!1)}\,.
$$
Note that
$$
\psi_{2k}^0(0)=\frac{H_{2k}(0)}{(\sqrt{\pi}\,2^{2k}(2k)!)^{1/2}}=
\frac{(-1)^k2^k(2k\!-\!1)!!}{(\sqrt{\pi}\,2^{2k}(2k)!)^{1/2}}=
\frac{(-1)^k}{\pi^{1/4}}\,\sqrt{E_k}\,,
$$
$$
(\psi_{2m+1}^0)'(0)=\frac{H_{2m+1}'(0)}{(\sqrt{\pi}\,2^{2m+1}(2m\!+\!1)!)^{1/2}}=
\frac{(-1)^m2^{m+1}(2m\!+\!1)!!}{(\sqrt{\pi}\,2^{2m+1}(2m\!+\!1)!)^{1/2}}=
\frac{(-1)^m\sqrt{2}}{\pi^{1/4}}\,\sqrt{(2m\!+\!1)E_m}\,.
$$
Therefore,
$$
(\widetilde{\p}_{2m+1}^0\,,\widetilde{\psi}_{2k}^0)_{+}= ({\p}_{2m+1}^0\,,{\psi}_{2k}^0)_{+}=
\frac{1}{\sqrt{2\pi}}\cdot\frac{(-1)^{m-k}}{2(m\!-\!k)\!+\!1}\,\sqrt{(2m\!+\!1)E_mE_k}\,.
$$
Since $\|\widetilde{\psi}_{2k}^0\|_{+}^2\!=\!\frac{1}{2}$\,, we obtain
$$
\frac{\widetilde{\psi}_{2m+1}^0}{\sqrt{(2m\!+\!1)E_m}}=
\sqrt{\frac{2}{\pi}}\,\sum_{k=0}^{+\infty} \frac{(-1)^{m-k}}{2(m\!-\!k)\!+\!1}\cdot
\sqrt{E_k}\,\widetilde{\psi}_{2k}^0\,.
$$
It gives
$$
P_+\biggl(\frac{(F^+q)(\zeta)}{\sqrt{\zeta}}\biggr)\equiv \frac{2^{3/4}}{\pi^{5/4}}\,
P_+\biggl(\sum_{l=-\infty}^{+\infty}\frac{(-1)^l}{2l\!+\!1}\,\zeta^l\cdot
\sum_{k=0}^{+\infty}\sqrt{E_k}(q,\widetilde\psi_{2k}^0)_{+}\zeta^k\biggr)
$$
$$
\equiv \frac{2^{1/4}}{\pi^{3/4}}\,\sum_{m=0}^{+\infty}
\frac{(q,\widetilde{\psi}_{2m+1}^0)_{+}}{\sqrt{(2m\!+\!1)E_m}}\,z^m \equiv
-\frac{2}{\pi}\,(G^+q)(z)\,,\quad z\!\in\!\D\,,
$$
where definition (\ref{FqGqDef}) of the functions $F^+q$ and $G^+q$ has been used.
\end{proof}
We should mention that Lemma \ref{*GasF} gives the simple proof of \cite{CKK} Theorem 4.2
about the equivalent definition of $\cH_0$\,. Following \cite{CKK}, we introduce the linear
operator
\begin{equation}
\label{OpSqrtZeta} (\cA f)(z)\equiv P_+\biggl[\frac{f(\zeta)}{\sqrt{-\zeta}}\biggr] \equiv
P_+\biggl[\sqrt{1\!-\!\overline{\zeta}}\cdot\frac{f(\zeta)}{\sqrt{1\!-\!\zeta}}\biggr]\,,
\qquad f\!\in\!L^1(\T)\,.
\end{equation}
Let
$$
\oo{H}^2_{3/4}=\{f\!\in\!\po{H}^2_{3/4}:f(1)\!=\!0\}\subset \po{H}^2_{3/4}\,.
$$
\begin{corollary}\label{*H0EquivDef} (i) The operator $\cA:\oo H^2_{3/4}\to\po H^2_{3/4}$ and
its inverse are bounded.\\
(ii) The following equality is fulfilled:
\begin{equation}
\label{cH0Def2} \cH_0=\biggl\{\{h_n\}_{n=0}^\infty: \sum_{n\ge 0}h_nz^n\equiv
P_+\biggl[\frac{g(\zeta)}{\sqrt{1\!-\!\overline{\zeta}}}\biggr],\ g\!\in\!H^2_{3/4}\biggr\}\,.
\end{equation}
The norms $\|h\|_{\cH}$ and $\|g\|_{H^2_{3/4}}$ are equivalent, i.e. $C_1\|g\|_{H^2_{3/4}}\le
\|h\|_{\cH}\le C_2\|g\|_{H^2_{3/4}}$ for any $g\in H^2_{3/4}$ and some absolute constants
$C_1\,,C_2>0$. \remark {\rm This equivalence was proved in \cite{CKK} using different and
complicated arguments.}
\end{corollary}
\begin{proof} (i) Recall that the mapping
$q\mapsto (G^+q)(-z)$ is a linear isomorphism between $\bH_+^0$ and $H^2_{3/4}$\,. Also, due
to the identity $(F^+q)(-1)\!=\!2^{-3/2}q(0)$ (see Proposition \ref{*NcNhasFG}), the mapping
$q\mapsto (F^+q)(-z)$ is a linear isomorphism between $\bH_+^0$ and $\oo{H}^2_{3/4}$\,.
Therefore, the mapping
$$
f(z)\equiv(F^+q)(-z)\mapsto q \mapsto (G^+q)(-z)\equiv
-\frac{\pi}{2}\,P_+\biggl[\frac{(F^+q)(-\zeta)}{\sqrt{\zeta}}\biggr]\equiv \cA f(z)
$$
is a linear isomorphism between $\oo H^2_{3/4}$\,, $\po \bH_+^0$ and $\po H^2_{3/4}$
respectively.\\
(ii) If $g\!\in\!H^2_{3/4}$\,, then $g_0(z)\!\equiv\!g(z)\!-\!g(1)\!\in\!\oo H^2_{3/4}$ and so
$|g_0(\zeta)|\!\le\!C|\zeta\!-\!1|^{1/4}$, $|\zeta|\!=\!1$\,, for some constant $C\!>\!0$\,.
Hence, the following equivalence is valid:
$$
\begin{array}{c}\displaystyle
\sum_{n\ge 0}h_nz^n\equiv P_{+}\biggl(\frac{g(\zeta)}{\sqrt{1\!-\!\overline{\zeta}}}\biggr) \
\ \Leftrightarrow\ \ \sum_{n\ge 0}h_nz^n\equiv g(1)+\frac{g_0(z)}{\sqrt{1\!-\!\overline{z}}}-
P_{-}\biggl(\frac{g_0(\zeta)}{\sqrt{1\!-\!\overline{\zeta}}}\biggr)
\cr\displaystyle\Leftrightarrow\ \ P_{+}\biggl(\sqrt{1\!-\!\overline{\zeta}}\sum_{n\ge 0}
h_n\zeta^n\biggr) {\vphantom{\Big|^\big|}}\equiv g(1)+g_0(z)\equiv g(z)\,,
\end{array}
$$
where $P_{-}f\!\equiv\!f\!-\!P_{+}f$ is the projector to the subspace of antianalytic
functions in $\D$\,. Therefore, the equation
$$
\frac{f(z)}{\sqrt{1\!-\!z}}\equiv\sum_{n\ge 0}h_nz^n\equiv
P_+\biggl[\frac{g(\zeta)}{\sqrt{1\!-\!\overline{\zeta}}}\biggr]\,,\quad {\rm where}\quad
f\!\in\!\oo{H}^2_{3/4}\,,\ \ g\!\in\!\po{H}^2_{3/4}\,,
$$
is equivalent to $g(z)\equiv(\cA f)(z)$\,. Then (i) follows (\ref{cH0Def2}).
\end{proof}

Consider the linear terms in asymptotics (\ref{LnSnAsympt}). Due to Lemma \ref{*cHEvenOdd},
the identities
$$
\sum_{n\ge 0} \nh q_{2n}^+z^n\equiv\frac{(F_Nq)(z)}{\sqrt{1\!-\!z}}\,, \ \quad {\rm where}
\quad (F_Nq)(z^2)\equiv\frac{(F^+q)(z)\sqrt{1\!+\!z}+(F^+q)(-z)\sqrt{1\!-\!z}}{2}\,,
$$
$$
\sum_{n\ge 0} \nh q_{2n+1}^+z^n \equiv \frac{(F_Dq)(z)}{\sqrt{1\!-\!z}}\,,\quad {\rm
where}\quad (F_Dq)(z^2)\equiv\frac{(F^+q)(z)\sqrt{1\!+\!z}-(F^+q)(-z)\sqrt{1\!-\!z}}{2z}
$$
hold true. Note that
\begin{equation}
\label{xxF} (F^+q)(z)\equiv \frac{(F_Nq)(z^2)\!+\!z(F_Dq)(z^2)}{\sqrt{1\!+\!z}}\,, \quad
z\!\in\!\D\,.
\end{equation}
\begin{lemma}
\label{*G+AsF+} For each $q\!\in\!\bH_+$ there exists functions $G_Nq\,,G_Dq\in H^2_{3/4}$
such that the following identities are fulfilled:
$$
\sum_{n\ge 0} \nc q_{2n}^+z^n\equiv P_+\biggl[
\frac{(G_Nq)(\zeta)}{\sqrt{1\!-\!\overline{\zeta}}}\biggr]\,, \qquad  \sum_{n\ge 0} \nc
q_{2n+1}^+z^n \equiv P_+\biggl[ \frac{(G_Dq)(\zeta)}{\sqrt{1\!-\!\overline{\zeta}}}\biggr]\,.
$$
Moreover,
$$
(G_Nq)(z)\equiv -\frac{\pi}{2}\,(F_Dq)(z)\,,\quad (G_Dq)(z)\equiv
-\frac{\pi}{2}\,(S_*F_Nq)(z)\,,\quad z\!\in\!\D\,,
$$
where $(S_*F)(z)\equiv P_+[\,\overline{\zeta}\cdot F(\zeta)]$ is the "left-shift"\ operator.
\end{lemma}
\begin{proof}
Define the function
$$
(G_Nq)(z)\equiv P_+\biggl[\sqrt{1\!-\!\overline{\zeta}} \sum_{n\ge 0} \nc
q_{2n}^+\zeta^{n}\biggr]\,\ \ \ \ z\in \D.
$$
It gives
$$
P_+\biggl[\frac{(G_Nq)(\zeta)}{\sqrt{1\!-\!\overline{\zeta}}}\biggr]\equiv
P_+\biggl[\,\sum_{n\ge 0} \nc q_{2n}^+\zeta^{n}- \frac{P_-[\sqrt{1\!-\!\overline{\zeta}}
\sum_{n\ge 0} \nc q_{2n}^+\zeta^{n}]}{\sqrt{1\!-\!\overline{\zeta}}}\,\biggr]\equiv
\sum_{n\ge 0} \nc q_{2n}^+z^{n}\,.
$$
On the other hand, we have
$$
\sum_{n\ge 0} \nc q_{2n}^+z^{2n}\equiv \frac{1}{2}\,\biggl(
P_+\biggl[\frac{(G^+q)(\zeta)}{\sqrt{1\!-\!\overline{\zeta}}}\biggr]+
P_+\biggl[\frac{(G^+q)(-\zeta)}{\sqrt{1\!+\!\overline{\zeta}}}\biggr]\biggr).
$$
Therefore,
$$
(G_Nq)(z^2)\equiv \frac{1}{2}\, P_+\biggl[{\sqrt{1\!+\!\overline{\zeta}}}\cdot(G^+q)(\zeta)+
{\sqrt{1\!-\!\overline{\zeta}}}\cdot(G^+q)(-\zeta)\biggr]\,.
$$
Lemma \ref{*GasF} yields
$$
(G_Nq)(z^2)\equiv  -\frac{\pi}{4}\,
P_+\biggl[\frac{\sqrt{1\!+\!\overline{\zeta}}}{\sqrt{\zeta}}\cdot(F^+q)(\zeta)+
\frac{\sqrt{1\!-\!\overline{\zeta}}}{\sqrt{-\zeta}}\cdot(F^+q)(-\zeta)\biggr]\,.
$$
Due to (\ref{xxF}), we obtain
$$
(G_Nq)(z^2)\equiv -\frac{\pi}{4}\,P_+\!\left[
\biggl(\frac{\sqrt{1\!+\!\overline{\zeta}}}{\sqrt\zeta\sqrt{1\!+\!\zeta}}+
\frac{\sqrt{1\!-\!\overline{\zeta}}}{\sqrt{-\zeta}\sqrt{1\!-\!\zeta}}\biggr)(F_Nq)(\zeta^2)
\right.
$$
$$
\left.+\biggl(\frac{\zeta\sqrt{1\!+\!\overline{\zeta}}}{\sqrt\zeta\sqrt{1\!+\!\zeta}}-
\frac{\zeta\sqrt{1\!-\!\overline{\zeta}}}{\sqrt{-\zeta}\sqrt{1\!-\!\zeta}}\biggr)(F_Dq)(\zeta^2)
\right]\equiv -\frac{\pi}{2}\,(F_Dq)(z^2)\,,
$$
where we have used the identities
$$
\frac{\sqrt{1\!+\!\overline{\zeta}}}{\sqrt{\zeta}\sqrt{1\!+\!\zeta}}= \frac{1}{\zeta}\,,\ \
\zeta\!\ne\!-1\,,\quad {\rm and}\quad
\frac{\sqrt{1\!-\!\overline{\zeta}}}{\sqrt{-\zeta}\sqrt{1\!-\!\zeta}}= -\frac{1}{\zeta}\,,\ \
\zeta\!\ne\!1\,.
$$
The proof of the identity
$(G_Dq)(z)\!\equiv\!-\frac{\pi}{2}\,P_+\left[\,\overline{\zeta}\cdot(F_Nq)(\zeta)\right]$ is
similar.
\end{proof}

\begin{proof}[{\bf Proof of Theorem \ref{*H+DChar} (i).\ }]
Recall that Theorem \ref{*Asympt} gives $\widetilde{s}_{2n+1}(q)=\nc q_{2n+1}^+ +
\el2_{\frac{3}{4}+\delta}(n)$, which yields
$$
\sum_{n\ge 0}\widetilde{s}_{2n+1}(q)z^n\equiv\sum_{n\ge 0}\nc q_{2n+1}^+z^n +
\omega(z)\,,\quad \omega\!\in\!H^2_{\frac{3}{4}+\delta}\,,\ \ \delta\!>\!0\,.
$$
Lemma \ref{*G+AsF+} implies
$$
\sum_{n\ge 0}\widetilde{s}_{2n+1}(q)z^n \equiv
P_+\biggl[\frac{(G_Dq)(\zeta)}{\sqrt{1\!-\!\overline{\zeta}}}\biggr]+\omega(z)\equiv
-\frac{\pi}{2}\, P_+\biggl[\frac{(F_Nq)(\zeta)}{\zeta\sqrt{1\!-\!\overline{\zeta}}}\biggr]
+\omega(z)\,.
$$
Since $(F_Nq)(z)\equiv \sqrt{1\!-\!z}\sum_{m\ge 0}\nh q_{2m}^+z^m$ and
$\sqrt{1\!-\!\zeta}\big/\zeta\sqrt{1\!-\!\overline{\zeta}}=-1\big/\sqrt{-\zeta}$\,,
$\zeta\!\ne\!1$\,, we obtain
\begin{equation}
\label{xxS2n+1} \sum_{n\ge 0}\widetilde{s}_{2n+1}(q)z^n\equiv
\frac{\pi}{2}\,P_+\biggl[\frac{1}{\sqrt{-\zeta}}\sum_{m\ge 0}\nh q_{2m}^+\zeta^m\biggr]
+\omega(z)\,.
\end{equation}
Let
$$
(\dlt H q) (z)\equiv{\sum_{m\ge 0}}(\nh q_{2m}^+\!-\!\nh q_{2m+1}^+)z^m\,,\quad z\!\in\!\D\,.
$$
Note that Lemma \ref{*cHEvenOdd} yields $\dlt H q\in H^2_{3/4}$
and Theorem \ref{*Asympt} gives
$$
\sum_{m\ge 0}\nh q_{2m}^+z^m\equiv (\dlt H q)(z)+\sum_{m\ge 0}\nh q_{2m+1}^+z^m\equiv (\dlt H
q)(z)+ \frac{1}{2}\sum_{m\ge 0}\mu_{2m+1}(q)z^m + \omega_1(z)\,,\quad
\omega_1\!\in\!H^2_{\frac{3}{4}+\delta}\,.
$$
Moreover, (\ref{TraceNonLin}) yields $\omega_1\!\in\!\oo H^2_{\frac{3}{4}+\delta}$\,.
Substituting the last identity into (\ref{xxS2n+1}), we have
$$
\sum_{n\ge 0}\widetilde{s}_{2n+1}(q)z^n \equiv
\frac{\pi}{4}\,P_+\biggl[\frac{1}{\sqrt{-\zeta}}\sum_{m\ge
0}\mu_{2m+1}(q)\zeta^m\biggr]+\frac{\pi}{2}\,P_+\biggl[\frac{(\dlt
Hq)(\zeta)}{\sqrt{-\zeta}}\biggr]+\omega_2(z)\,,
$$
where $\omega_2(z)\!=\!\omega(z)\!+\!
\frac{\pi}{2}\,P_+\left[\omega_1(\zeta)\big/\sqrt{-\zeta}\,\right]\!\in\!
H^2_{\frac{3}{4}+\delta}$ since the operator $\po \cA: \oo H^2_{\frac{3}{4}+\delta}\to \po
H^2_{\frac{3}{4}+\delta}$ given by (\ref{OpSqrtZeta}) is bounded\begin{footnote}{The proof
repeats the proof of \cite{CKK} Theorem 4.2 or Corollary \ref{*H0EquivDef} in the present
paper.}\end{footnote}.

\pagebreak

Furthermore, denote values $\nt q_n^+$ by
\begin{equation}
\label{TildeQDef}
\sum_{n\ge 0}\nt q^+_nz^n
\equiv \frac{\pi}{2}\,P_+\biggl[\frac{(\dlt
Hq)(\zeta)\!-\!(\dlt Hq)(1)}{\sqrt{-\zeta}}\biggr]\equiv \frac{\pi}{2}\,P_+\biggl[\frac{(\dlt
Hq)(\zeta)}{\sqrt{-\zeta}}\biggr]-\frac{\pi
q(0)}{8}\,P_+\biggl[\frac{1}{\sqrt{-\zeta}}\biggr]\,,
\end{equation}
where we have used (\ref{Fq(-1)}).  Note that
$\{\nt{q}_n^+\}_{n=0}^{\infty}\!\in\!\el2_{3/4}$\,, since the operator $\po \cA:\oo
H^2_{3/4}\to \po H^2_{3/4}$ is bounded (see Corollary \ref{*H0EquivDef}). Summarizing, we
obtain
$$
\sum_{n\ge 0}\widetilde{s}_{2n+1}(q)z^n \equiv
\frac{\pi}{4}\,P_+\biggl[\frac{1}{\sqrt{-\zeta}}\sum_{m\ge 0}\mu_{2m+1}(q)\zeta^m\biggr] +
\frac{\pi q(0)}{8}\,P_+\biggl[\frac{1}{\sqrt{-\zeta}}\biggr] + \sum_{n\ge 0}\nt q_n^+z^n
+\omega_2(z)\,.
$$
Since $1\big/\sqrt{-\zeta}=\frac{2}{\pi}\sum_{l=-\infty}^{+\infty}\zeta^l\big/(2l\!+\!1)$\,,
it implies
\begin{equation}
\label{TildeQAsymp} \widetilde{s}_{2n+1}(q)= \frac{1}{2}\sum_{m\ge 0}
\frac{\mu_{2m+1}(q)}{2(n\!-\!m)\!+\!1}+ \frac{q(0)}{4(2n\!+\!1)}+r_{2n+1}(q)\,,\qquad
r_{2n+1}(q)\!=\!\nt q^+_n+ \el2_{\frac{3}{4}+\delta}(n)\,.
\end{equation}
In particular, $\{r_{2n+1}(q)\}_{n=0}^\infty\in\el2_{3/4}$\,.
\end{proof}
\remark Due to definition of $F^+q$ and Lemma \ref{*cHEvenOdd}, the mappings $q\mapsto
F^+q\mapsto (F_Dq\,,\dlt H q)$ are linear isomorphisms between $\bH_+$\,, $H^2_{3/4}$ and
$H^2_{3/4}\times H^2_{3/4}$ respectively. Since the operator
$\cA:\oo{H}^2_{3/4}\!\to\!\po{H}^2_{3/4}$ and its inverse are bounded (see Corollary
\ref{*H0EquivDef}) and due to definition (\ref{TildeQDef}) of $\nt q_n^+$, the mappings
$$
\dlt Hq \mapsto \left((\dlt Hq)(0);\dlt H q\!-\!(\dlt Hq)(0)\right)\mapsto \left((\dlt
Hq)(0);{\textstyle \sum_{n\ge 0}\nt q_n^+z^n}\right)\,,
$$
are linear isomorphisms between $\po H^2_{3/4}$\,, $\po \R\times \oo H^2_{3/4}$ and $\po \R
\times \po H^2_{3/4}$\,. Therefore, using the identity $(\dlt Hq)(0)\!=\!\frac{1}{4}\,q(0)$\,,
we obtain that the mappings
\begin{equation}
\label{TildeQIso}
\begin{array}{c}\displaystyle
q\mapsto (F_Dq\,;\dlt H q)
\mapsto \left(\{\nh q_{2n+1}^+\}_{n=0}^\infty\,;q(0)\,;\{\nt
q_n^+\}_{n=0}^\infty\right)\,,\cr \displaystyle \bH_+\ \leftrightarrow\ \po
H^2_{3/4}\times \po H^2_{3/4} \quad \leftrightarrow
\quad\quad \cH\times\R\times\el2_{3/4}\,,\qquad\quad
\end{array}
\end{equation}
are linear isomorphisms.

\vskip 4pt \noindent {\bf Proof of Theorem \ref{*H+DChar} (ii).\ } Consider the mapping
$\Psi:\bH_+\to\cS_D\times\R\times\el2_{3/4}$\,, given by
$$
\Psi: q\mapsto \biggl(\{\mu_{2n+1}(q)\}_{n=0}^\infty\,,q(0)\,,
\{r_{2n+1}(q)\}_{n=0}^\infty\biggr)\,,\qquad
$$
which maps potential to spectral data. The proof given below consists of four steps:\linebreak
1) $\Psi$ is injective; 2) $\Psi$ is analytic; 3) $\Psi$ is a local isomorphism; 4) $\Psi$ is
surjective.

\begin{proof}[{\bf 1. Uniqueness Theorem.}]
The proof repeats the case of the whole real line \cite{CKK1}. Let
$$
\left(\{\mu_{2n+1}(p)\}_{n=0}^\infty\,,p(0)\,, \{r_{2n+1}(p)\}_{n=0}^\infty\right)=
\left(\{\mu_{2n+1}(q)\}_{n=0}^\infty\,,q(0)\,, \{r_{2n+1}(q)\}_{n=0}^\infty\right)\,,
$$
for some potentials $p,q\!\in\!\bH_+$. Note that it gives
$$
\l_{2n+1}=\l_{2n+1}(p)\!=\!\l_{2n+1}(q)\,,\quad s_{2n+1}=s_{2n+1}(p)\!=\!s_{2n+1}(q)\,,\quad
n\!\ge\!0\,.
$$
Recall that $\varphi(x,\l,q)$ is the solution of Eq. (\ref{PertEq}) such that
$\vp(0,\l,q)\!=\!0$\,, $\vp'(0,\l,q)\!=\!1$\,. Define functions
$$
f_1(\l;x,q,p)=\frac{F_1(\l;x,q,p)}{w_D(\l,q)}\,,\ \
F_1(\l;x,q,p)=\varphi(x,\l,p)\psi'_+(x,\l,q)- \psi_+(x,\l,p)\varphi'(x,\l,q)\,,
$$
$$
f_2(\l;x,q,p)=\frac{F_2(\l;x,q,p)}{w_D(\l,q)}\,,\ \
F_2(\l;x,q,p)=\varphi(x,\l,p)\psi_+(x,\l,q)- \psi_+(x,\l,p)\varphi(x,\l,q)\,,
$$
where
$$
w_D(\l,q)=\{\vp(\cdot,\l,q)\,,\psi_+(\cdot,\l,q)\}= -\psi_+(0,\l,q)\,.
$$

Both functions $f_1$\,, $f_2$ are entire with respect to $\l$. Indeed, all roots
$\l_{2n+1}(p)\!=\!\l_{2n+1}(q)$ of the denominator $w_D(\cdot,q)$ are simple and all these
values are roots of the numerators $F_1$\,, $F_2$\,, since the  definition of norming
constants yields
$$
\varphi(x,\l_{2n+1}\,,q)\!= e^{s_{2n+1}}\cdot\psi_+(x,\l_{2n+1}\,,q),\quad
\varphi(x,\l_{2n+1}\,,p)\!= e^{s_{2n+1}}\cdot\psi_+(x,\l_{2n+1}\,,p)\,.
$$
Standard estimates (see Lemma \ref{*AEstim} and asymptotics (\ref{EstimExact})) of $\varphi$
and $\psi_+$ give
$$
f_1(\l;x,p,q)=O(1)\,,\quad f_2(\l;x,p,q)=O(|\l|^{-1/2})\,, \quad
|\l|\!=\!\l_{2n'}^0\,,\ n'\!\to\!\infty\,.
$$
Furthermore, the maximum principle and asymptotics of $f_1$ as $\l\!\to\!-\infty$ implies
$$
f_1(\l;x,p,q)= 1\,,\quad f_2(\l;x,p,q)= 0\,, \quad\l\!\in\!\C\,.
$$
This follows $\vp(x,\l,p)=\vp(x,\l,q)$ and $\psi_+(x,\l,p)=\p_+(x,\l,q)$, i.e. $p=q$\,.
\end{proof}

\begin{proof}[{\bf 2. Analyticity of $\Psi$.}]
Recall that for some $\delta\!>\!0$ the following asymptotics are fulfilled (see \nolinebreak
(\ref{LnSnAsympt}) and (\ref{TildeQAsymp})):
\begin{equation}
\label{MuSAsymptD} \mu_{2n+1}(q)=2\nh q_{2n+1}^+ + \el2_{\frac{3}{4}+\delta}(n)\,,\qquad
r_{2n+1}(q)=\nt q_n^+ + \el2_{\frac{3}{4}+\delta}(n)\,.
\end{equation}
Let $\bH_{+\C}$ be the complexification of the space $\bH_+$\,. Due to Lemma
\ref{*AAnalyticity}, for each real potential $q\!\in\!\bH_+$ all spectral data
$\mu_{2n+1}(q)$\,, $s_{2n+1}(q)$\,, $n\!\ge\!0$\,, extend analytically to some complex
neighborhood ${Q}\!\subset\!\bH_{+\C}$ of $q$\,. Repeating the proof of (\ref{MuSAsymptD}) we
obtain that these asymptotics hold true uniformly on bounded subsets of ${Q}$\,. Note that
Lemma \ref{*AAnalyticity} gives
\begin{equation}
\label{GradIdent} \frac{\partial\mu_{2n+1}(q)}{\partial q(t)}= \psi_{2n+1}^2(t,q)\,,\qquad
\frac{\partial s_{2n+1}(q)}{\partial q(t)}= (\psi_{2n+1}\chi_{2n+1})(t,q)\,,
\end{equation}
where $\psi_{2n+1}(\cdot,q)$ is the $n$-th normalized eigenfunction of $T_D$ and
$\chi_{2n+1}(\cdot,q)$ is some special solution of Eq. (\ref{PertEq}) for
$\l\!=\!\l_{2n+1}(q)$ such that $\{\chi_{2n+1}\,,\psi_{2n+1}\}\!=\!1$\,. In particular,
$$
(\psi_{2n+1}\chi_{2n+1})(t,q)\sim t\,,\ \ t\!\to\!0\,,\qquad (\psi_{2n+1}\chi_{2n+1})(t,q)\sim
-t^{-1},\ \ t\!\to\!+\infty\,.
$$

\pagebreak

Let
$$
\Psi = \Psi_0 + \Psi_1\,,\quad \Psi_0:q\mapsto\left(\{2\nh
q_{2n+1}^+\}_{n=0}^\infty\,; q(0)\,; \{\nt q_n^+\}_{n=0}^\infty
\right)\,, \quad \Psi_1: q \mapsto\Psi q-\Psi_0 q\,.
$$
As it was shown above (see (\ref{TildeQIso})), the linear operator $\Psi_0$ is an isomorphism
between $\bH_+$ and $\cH\times\R\times\el2_{3/4}$\,. In particular, $\Psi_0$ is a
real-analytic mapping. Consider the second term
$$
\Psi_1:q\mapsto \biggl(\{\mu_{2n+1}(q)\!-\!2\nh q_{2n+1}^+\}_{n=0}^\infty\,; 0\,;
\{r_{2n+1}(q)\!-\!\nt q_n^+\}_{n=0}^\infty\biggr)\,,
$$
$$
\Psi_1:\bH_+\to\el2_{\frac{3}{4}+\delta}\times\R\times\el2_{\frac{3}{4}+\delta} \subset
\cH\times\R\times\el2_{\frac{3}{4}}\,.
$$
Each "coordinate function"\ $\mu_{2n+1}(q)\!-\!2\nh q_{2n+1}^+$\,, $r_{2n+1}(q)\!-\!\nt q_n^+$
is analytic and the mapping $\Psi_1$ is correctly defined and bounded in some complex
neighborhood of each real potential (since (\ref{MuSAsymptD}) holds true uniformly on bounded
subsets). Then, $\Psi_1$ is a real-analytic mapping from $\bH_+$ to
$\el2_{\frac{3}{4}+\delta}\times\R\times\el2_{\frac{3}{4}+\delta}$\,. Hence,
$\Psi=\Psi_0+\Psi_1$ is a real-analytic mapping.
\end{proof}

\begin{proof}[{\bf 3. Local Isomorphism.}] Since $\Psi$ is analytic, the Fr\'echet derivative
$d_q\Psi:\bH_+\!\to\!\cH\times\R\times\el2_{3/4}$ is bounded for each $q\!\in\!\bH_+$\,. We
need to show that $(d_q\Psi)^{-1}$ is bounded too. Note that the operator
$$
d_q\Psi_1:\bH_+\to \cH\times\R\times\el2_{\frac{3}{4}}
$$
is compact since it maps $\bH_+$ into
$\el2_{\frac{3}{4}+\delta}\times\R\times\el2_{\frac{3}{4}+\delta}$ and the embedding
$\el2_{\frac{3}{4}+\delta}\subset \el2_{\frac{3}{4}}$ is compact. Therefore,
$$
d_q\Psi=d_q\Psi_0+d_q\Psi_1=\Psi_0+d_q\Psi_1: \bH_+\to \cH\times\R\times\el2_{\frac{3}{4}}
$$
is a Fredholm operator (due to (\ref{TildeQIso}), the inverse operator $\Psi_0^{-1}$ is
bounded). By Fredholm's Theory, in order to prove that $(d_q\Psi)^{-1}$ is bounded, it is
sufficient to check that the range $\Ran d_q\Psi$ is dense:
\begin{equation}
\label{ClImPsi} \cH\times\R\times\el2_{3/4}=\overline{\Ran d_q\Psi}\,.
\end{equation}

Due to Lemma \ref{*AGradProd} (i), for each $q\in\bH_+$ the following standard identities are
fulfilled:
\begin{equation} \label{GradProd}
\begin{array}{c}\displaystyle
\left((\psi_{2n+1}^2)'(q),\psi_{2m+1}^2(q)\right)_{+}=0\,,\quad
\left((\psi_{2n+1}^2)'(q),(\psi_{2m+1}\chi_{2m+1})(q)\right)_{+}=
{\textstyle\frac{1}{2}}\,\delta_{mn}\,, \vphantom{\big|_{\big|}}\cr \displaystyle
\vphantom{\big|^{\big|}}
\left((\psi_{2n+1}\chi_{2n+1})'(q),(\psi_{2m+1}\chi_{2m+1})(q)\right)_{+}=0\,,\quad
n,m\!\ge\!0.
\end{array}
\end{equation}
Using (\ref{GradIdent}) and (\ref{GradProd}), we obtain
$$
\biggl(2(\psi_{2n+1}^2)'(q)\,, \frac{\partial\mu_{2m+1}(q)}{\partial
q}\biggr)_{\!+}\!\!=0\,,\quad \biggl(2(\psi_{2n+1}^2)'(q)\,, \frac{\partial\widetilde
s_{2m+1}(q)}{\partial q}\biggr)_{\!+}\!\!=\delta_{nm}\,.
$$
Note that $(\psi_{2n+1}^2)'(\cdot,q)\!\in\!\bH_+$ and $(\psi_{2n+1}^2)'(0,q)\!=\!0$\,. By
definition (\ref{RnDefD}) of $r_{2m+1}(q)$\,, we have
$$
\biggl(2(\psi_{2n+1}^2)'(q)\,, \frac{\partial r_{2m+1}(q)}{\partial
q}\biggr)_{\!+}\!\!=\delta_{nm}\,.
$$
Hence,
$$
(d_q\Psi)\left(2(\psi_{2n+1}^2)'(\cdot,q)\right) = \left({\bf 0}\,; 0\,;{\bf
e}_n\right)\,,
$$
where ${\bf 0}\!=\!(0,0,0,\dots)$\,, ${\bf e}_0\!=\! (1,0,0,\dots)$\,, ${\bf
e}_1\!=\!(0,1,0,\dots)$ and so on. Therefore,
\begin{equation}
\label{xxClImPsi} \left\{({\bf 0};0)\right\}\times\el2_{3/4} =\{({\bf 0};0;{\bf c}):{\bf
c}\!\in\!\el2_{3/4}\} \subset\overline{\Ran d_q\Psi}\,.
\end{equation}

We come to the second component of $(d_q\Psi)\xi$\,, i.e. to the value $\xi(0)$\,.
We consider the lowest eigenvalue $\l_0(q)$ of the operator $T_N$ (with the same potential $q$)
and the function
$$
\xi(t)= (\varphi\vartheta)'(t,\l_0(q),q)\,,\quad t\in \R\,.
$$
Note that $\xi(0)\!=\!1$ and asymptotics (\ref{Psi+Asympt1}), (\ref{Chi+Asympt1}) give
$\xi\!\in\!\bH_+$ since $\vartheta(\cdot,\l_0(q),q)$ is proportional to
$\psi_+(\cdot,\l_0(q),q)$\,. Moreover, since
$\varphi(0,\lambda_0(q),q)\!=\!\psi_{2m+1}(0,q)\!=\!0$\,, we have
$$
\biggl(\xi\,, \frac{\partial\mu_{2m+1}(q)}{\partial
q}\biggr)_{\!+}\!\!=\left(\xi,\psi_{2m+1}^2(q)\right)_{+}=
\frac{-\{\psi_{2m+1}\,,\varphi\}\{\psi_{2m+1}\,,\vartheta\}(0,\l_0(q),q)}
{2(\l_{2m+1}(q)\!-\!\l_0(q))}=0\,.
$$
Hence,
$$
(d_q\Psi)\xi = \left({\bf 0}\,; 1\,; (d_q r)\xi \right)\,,\quad  {\rm where}\quad (d_q r)\xi=
\biggl\{\biggl(\xi\,,\frac{\partial r_{2n+1}(q)}{\partial q}\biggr)_{\!+}\biggr
\}_{n=0}^{\infty}\in\el2_{3/4}\,.
$$
Together with (\ref{xxClImPsi}) this gives
$$
\{{\bf 0}\}\times\R\times\el2_{3/4}\subset\overline{\Ran d_q\Psi}\,.
$$

Furthermore, we consider the functions $-2(\psi_{2m+1}\chi_{2m+1})'(q)\in\bH_+$ (see
asymptotics (\ref{Psi+Asympt1}), (\ref{Chi+Asympt1})). Identities (\ref{GradIdent}),
(\ref{GradProd}) and $(\psi_{2m+1}\chi_{2m+1})'(0,q)\!=\!1$ give
$$
(d_q\Psi)\left(-2(\psi_{2m+1}\chi_{2m+1})'(q)\right) = \left({\bf e}_m\,;
-2\,;(d_qr)\left(-2(\psi_{2m+1}\chi_{2m+1})'(q)\right)
\right)\,.
$$
Due to Proposition \ref{*H0Emb} (ii), the set of finite sequences is dense in $\cH_0$\,.
Therefore,
\begin{equation}
\label{xxxxClImPsi} \cH_0\times\R\times\el2_{3/4}\subset\overline{\Ran d_q\Psi}\,.
\end{equation}
In conclusion, we consider an arbitrary function $\zeta\!\in\!\bH_+$ such that
$\int_{\R_+}\!\zeta(t)dt\!\ne\!0$. Proposition \ref{*H0Emb} (i) follows
$$
(d_q\Psi)\zeta\notin\cH_0\times\R\times\el2_{3/4}\,.
$$
Together with (\ref{xxxxClImPsi}) it yields (\ref{ClImPsi}), since the codimension of $\cH_0$
in $\cH$ is equal to $1$.
\end{proof}

\begin{proof}[{\bf 4. $\Psi$ is surjective.}] We need
\begin{lemma}
\label{*S2k+1Change} Let $q\!\in\!\bH_+$\,, $n\!\ge\!0$ and $t\!\in\!\R$\,. Denote
$$
q_{2n+1}^t(x)=q(x)-2\frac{d^2}{dx^2}\log\eta_{2n+1}^t(x,q)\,, \quad \eta_{2n+1}^t(x,q)=
1+(e^t\!-\!1)\int_x^{+\infty}\!\!\psi_{2n+1}^2(s,q)ds\,.
$$
Then $q_{2n+1}^t\!\in\!\bH_+$ and
$$
\l_{2m+1}(q_{2n+1}^t)=\l_{2m+1}(q)\,, \qquad s_{2m+1}(q_{2n+1}^t)=s_{2m+1}(q)+t\delta_{mn}
$$
for all $m\!\ge\!0$\,. Moreover, $q_{2n+1}^t(0)\!=\!q(0)$\,. \remark {\rm By definition
(\ref{RnDefD}), we have $r_{2m+1}(q_n^t)\!=\!r_{2m+1}(q)\!+\!t\delta_{nm}$ for all
$m\!\ge\!0$\,.}
\end{lemma}
\begin{proof} This Lemma is similar to \cite{CKK} Theorem 3.5 and can be
proved by direct calculations using the so-called Darboux transform of second-order
differential equation (see also \cite{MT}, \cite{PT}). Note that
$$
\eta_{2n+1}^t(x,q)= e^t-(e^t\!-\!1)\int_0^x\!\!\psi_{2n+1}^2(s,q)ds=e^t+O(x^3)\,,\quad
x\downarrow 0\,.
$$
This implies $q_{2n+1}^t(0)=q(0)$\,.
\end{proof}

We consider an arbitrary spectral data $(h^*;u^*;c^*)\in\cS_D\times\R\times\el2_{3/4}$\,. Due
to Theorem \nolinebreak \ref{*H+2SpecChar}, there exists a potential $q^*\!\in\!\bH_+$ such
that
$$
\mu_{2n+1}(q^*)=h^*_n\,,\ \ n\!\ge\!0\,,\quad {\rm and}\quad q^*(0)=2{\textstyle \sum_{n\ge
0}}\tau_n(q^*)=u^*\,.
$$
It gives
$$
\left(h^*;u^*;r(q^*)\right)\in \Psi(\bH_+)\,,\quad {\rm where}\quad
r(q^*)\!=\!(r_1(q^*)\,,r_3(q^*)\,,\dots)\,.
$$
Due to Proposition \ref{*H0Emb} (ii), for each $\varepsilon\!>\!0$ there exist a finite
sequence $t_\varepsilon\!=\!(t_1\,,...\,,t_{2k-1}\,,0\,,...)$ such that
$$
\left\|(c^*\!-r(q^*))-t_\varepsilon\right\| = \left\|(c^*\!-t_\varepsilon)-r(q^*)\right\| <
\varepsilon\,.
$$
Since $\Psi$ is a local isomorphism, for some $\varepsilon\!>\!0$ we have
$$
\left(h^*;u^*;c^*\!-\!t_\varepsilon\right)=
\left(h^*;u^*;(c^*_1\!-\!t_1\,,\dots,c^*_{2k-1}\!-\!t_{2k-1}\,,
c^*_{2k+1}\,,c^*_{2k+3}\,,\dots)\right)\in \Psi(\bH_+)\,.
$$
It means that $\left(h^*;u^*;c^*\!-\!t_\varepsilon\right)\!=\!\Psi(q_k)$ for some
$q_k\!\in\!\bH_+$. Using Lemma \ref{*S2k+1Change} step by step, we construct the sequence of
potentials $q_j\!\in\!\bH_+$ such that
$$
q_{j-1}=(q_{j})_{2j-1}^{t_{2j-1}}\,,\quad j\!=\!k,k\!-\!1,\dots,1\,,
$$
$$
\Psi(q_j)=\left(h^*;u^*; (c^*_1\!-\!t_1\,,\dots,c^*_{2j-1}\!-\!t_{2j-1}\,, c^*_{2j+1}\,,
c^*_{2j+3}\,,\dots)\right)\,.
$$
Then, $\Psi(q_0)\!=\!(h^*;u^*;c^*)$\,.
\end{proof}

\section{Perturbed oscillator on $\R_+$, general case} \setcounter{equation}{0}

In this Section we consider the family of self-adjoint operators $\{T_b\}_{b\in\R}$ given by
$$
T_b y=-y''+x^2y+q(x)y\,,\quad y'(0)\!=\!b y(0)\,,\quad q\!\in\!\bH_+\,.
$$
We begin with asymptotics of spectral data. Recall that
$$
w_N(\l,q,b)=\p'_+(0,\l,q)-b\p_+(0,\l,q)\,,\quad \l\!\in\!\C\,.
$$
Due to standard estimates of $\psi_+$\,, $\psi'_+$ (see Lemma \ref{*AEstim} and asymptotics
(\ref{EstimExact})), we have
$$
w_N(\l,q,b)= (\psi_+^0)'(0,\l)\cdot(1\!+\!O(|\l|^{-1/2}))\,,\quad |\l|\!=\!\l_{2k+1}^0\,,\ \
k\!\to\!\infty\,.
$$
Repeating arguments of Lemma 4.2 \cite{CKK1} and using Rouch\'e's Theorem, we obtain
$$
\l_{2n}(q,b)=\l_{2n}^0+\mu_{2n}(q,b)\,,\quad \mu_{2n}(q,b)=O(n^{-1/2})\,,\quad
n\!\to\!\infty\,,
$$
uniformly on bounded subsets of $(q,b)\!\in\!\bH_+\!\times\!\R$\,.

We define the modified norming constant $\widetilde{s}_{2n}(q,b)$ by
\begin{equation}
\label{TildeSnDefNqb} s_{2n}(q,b)=s_{2n}^0+\alpha_{2n}\mu_{2n}(q,b)+\widetilde
s_{2n}(q,b)\,,\quad n\!\ge\!0\,,
\end{equation}
where $s_{2n}^0\!=\!s_{2n}(0,0)$ and
$\alpha_{2n}=-{\dot{\psi}_+^0}\big/{\psi_+^0}\,(0,\l_{2n}^0)$\,.
\begin{theorem}
\label{*AsymptN} For each $(q;b)\!\in\!\bH_+\times\R$ and some absolute constant
$\delta\!>\!0$ the following asymptotics are fulfilled:
\begin{equation}
\label{LSAsymptN} \mu_{2n}(q,b)=2{\nh q_{2n}^+} + \frac{2E_n}{\sqrt{\pi}}\,b+
\el2_{\frac{3}{4}+\delta}(n)\,,     \qquad \widetilde{s}_{2n}(q)={\nc q_{2n}^+} + \frac{\pi
E_n^2}{8}\,b^2 +\el2_{\frac{3}{4}+\delta}(n)\,,
\end{equation}
uniformly on bounded subsets of $\bH_+\!\times\!\R$\,. \remark {\rm Recall that
$E_n=2^{-2n}(2n)!\big/(n!)^2=\pi^{-\frac{1}{2}}n^{-\frac{1}{2}}(1\!+\!O(n^{-1}))$\,,
$n\!\to\!\infty$\,.}
\end{theorem}

\begin{proof} Let $\lambda_{2n}\!=\!\lambda_{2n}(q,b)$ and $\mu_{2n}\!=\!\mu_{2n}(q,b)$.
We use arguments similar to the proof of Theorem \nolinebreak \ref{*Asympt}. Due to Corollary
\ref{*ADotEstim}, Lemma \ref{*LA511} (i) and asymptotics (\ref{KappaValues}), we have
$$
\frac{\psi'_+(0,\l_{2n}^0\!+\!\mu_{2n}\,,q)}{\dot{\kappa}'_{2n}}= -2\nh q_{2n}^+ +
\mu_{2n}+O(n^{-1}\log^2n)\,,\quad \frac{\psi_+(0,\l_{2n}^0\!+\!\mu_{2n}\,,q)}{\kappa_{2n}}=
1+O(n^{-\frac{1}{2}}\log n)\,.
$$
Since $(\psi'_+\!-\!b\psi_+)(0,\l_{2n}^0\!+\!\mu_{2n}\,,q)\!=\!0$ and
$\kappa_{2n}=O((\lambda_{2n}^0)^{-\frac{1}{2}}\dot{\kappa}'_{2n})$ (see (\ref{KappaValues})),
we obtain
\begin{equation}
\label{MAsympt1} \mu_{2n}=2\nh q_{2n}^+ + \frac{\kappa_{2n}}{\dot{\kappa}'_{2n}}\,b +
O(n^{-1}\log^2 n)= 2\nh q_{2n}^+ + \frac{2E_n}{\sqrt{\pi}}\,b + O(n^{-1}\log^2 n)\,.
\end{equation}
Let $\mu_{2n}^{(1)}\!=\!2\nh q_{2n}^+ \!+\! {2\pi^{-\frac{1}{2}}E_n}b$\,. Using Lemma
\ref{*LA511} (i) again, we see that
$$
\frac{\psi'_+(0,\l_{2n}^0\!+\!\mu_{2n}\,,q)}{\dot{\kappa}'_{2n}}= -2\nh q_{2n}^+ +\mu_{2n} +
2\nc q_{2n}^+\nh q_{2n}^+ -\biggl(\frac{2\dot{\kappa}_{2n}}{\kappa_{2n}}\,\nh q_{2n}^+ \!+\!
\nc q_{2n}^+\biggr) \cdot \mu_{2n}^{(1)} +
\frac{\ddot{\kappa}'_{2n}}{2\dot{\kappa}'_{2n}}\,(\mu_{2n}^{(1)})^2+\el2_{\frac{3}{4}+\delta}(n)
$$
and
$$
\frac{\psi_+(0,\l_{2n}^0\!+\!\mu_{2n}\,,q)}{\kappa_{2n}} = 1 - \nc q_{2n}^+ +
\frac{\dot{\kappa}_{2n}}{\kappa_{2n}}\,\mu_{2n}^{(1)}+O(n^{-1}\log^2 n)\,.
$$
Substituting these asymptotics into the equation
$(\psi'_+\!-\!b\psi_+)(0,\l_{2n}^0\!+\!\mu_{2n}\,,q)\!=\!0$, we obtain
$$
\mu_{2n} = \mu_{2n}^{(1)} + \biggl(\frac{\dot{\kappa}_{2n}}{\kappa_{2n}}-
\frac{\ddot{\kappa}'_{2n}}{2\dot{\kappa}'_{2n}}\biggr)\cdot (\mu_{2n}^{(1)})^2 +
\el2_{\frac{3}{4}+\delta}(n)= \mu_{2n}^{(1)} + \el2_{\frac{3}{4}+\delta}(n)\,.
$$

We come to the second part of (\ref{LSAsymptN}). Due to Lemma \ref{*LA511} (i), we have
$$
e^{s_{2n}^0-s_{2n}(q,b)}=\frac{\psi_+(0,\l_{2n}\,,q)}{\kappa_{2n}}= 1-\nc q_{2n}^+ +
\frac{\dot{\kappa}_{2n}}{\kappa_{2n}}\,\mu_{2n}+ \frac{(\nc q_{2n}^+)^2}{2}\,- \frac{\pi^2(\nh
q_{2n}^+)^2}{8}
$$
$$
+ \biggl( - \frac{\dot{\kappa}_{2n}}{\kappa_{2n}}\,\nc q_{2n}^+ + \frac{\pi^2}{8}\,\nh
q_{2n}^+\biggr)\mu_{2n}^{(1)} + \frac{\ddot{\kappa}_{2n}}{2\kappa_{2n}}\,(\mu_{2n}^{(1)})^2 +
\el2_{\frac{3}{4}+\delta}(n)\,.
$$

\pagebreak Therefore,
$$
s_{2n}^0-s_{2n}(q,b)= -\nc q_{2n}^+ + \frac{\dot{\kappa}_{2n}}{\kappa_{2n}}\,\mu_{2n}  -
\frac{\pi^2}{8}\biggl(\nh q_{2n}^+ - \frac{\mu_{2n}^{(1)}}{2}\biggr)^{\!2} +
\biggl(\frac{\ddot{\kappa}_{2n}}{\kappa_{2n}}- \frac{\dot{\kappa}_{2n}^2}{\kappa_{2n}^2} +
\frac{\pi^2}{16}\biggr)\frac{(\mu_{2n}^{(1)})^2}{2}+ \el2_{\frac{3}{4}+\delta}(n)\,.
$$
$$
= -\nc q_{2n}^+ + \frac{\dot{\kappa}_{2n}}{\kappa_{2n}}\,\mu_{2n} - \frac{\pi E_n^2}{8}\,b^2 +
\el2_{\frac{3}{4}+\delta}(n)\,.
$$
This asymptotic is equivalent to (\ref{LSAsymptN}) due to definition (\ref{TildeSnDefNqb}) of
$\widetilde{s}_{2n}(q,b)$.
\end{proof}

\begin{lemma}
\label{*TraceNLN} Let $(q,b)\!\in\!\bH_+\times\R$\,. Then the following identity is fulfilled:
$$
\sum_{n\ge 0} (\mu_{2n}(q,b)\!-\!2\nh q_{2n}^+ - 2\pi^{-1/2}E_{n}b) = -\frac{b^2}{2}\,.
$$
\remark {\rm This formula generalizes the first identity in (\ref{TraceNonLin}).}
\end{lemma}
\begin{proof} We put $q(-x)\!=\!q(x)$\,, $x\!\ge\!0$\,, and consider the following symmetric
potential on $\R$\,:
$$
q_\delta(x)= q(x)+2b\cdot\delta(x)\,.
$$
Then  (formally) we have $\l_{2n}(q_\delta)\!=\!\l_{2n}(q,b)$\,. Note that
$$
(q_\delta)\vphantom{\big|}^{\wedge}_{2n}=2\nh q_{2n}^+ + (\psi_{2n}^0)^2(0)\cdot 2b = 2\nh
q_{2n}^+ + {2\pi^{-1/2}E_n}\cdot b\,.
$$
 The right-hand side coincides with the main (linear with respect to $q,b$) term in
the first asymptotics (\ref{LSAsymptN}). Repeating the proof of (\ref{TraceNonLin}) (and Lemma
\ref{*ALemma}), we obtain
$$
\sum_{n\ge 0} (\mu_{2n}(q,b)\!-\!2\nh q_{2n}^+ - 2\pi^{-1/2}E_{n}b)= 2\int_0^1 ds\int_0^s
\sum_{n\ge 0}\sum_{m:m\ne n}
\frac{\left(q_\delta,(\psi_{2n}\psi_{2m})(\cdot,tq_\delta)\right)_{L^2(\R)}^2}
{\l_{2n}(tq_\delta)\!-\!\l_{2m}(tq_\delta)}\,dt
$$
$$
=8b^2 \int_0^1 ds\int_0^s \sum_{n\ge 0}\sum_{m:m\ne n}
\frac{(\psi_{2n}\psi_{2m})^2(0,tq_\delta)} {\l_{2n}(tq_\delta)\!-\!\l_{2m}(tq_\delta)}\,dt\,.
$$
Since $(\psi_{2n}^0)^2(0)\!=\!\pi^{-1/2}E_n=\pi^{-1}(n\!+\!1)^{-1/2}+O(n^{-3/2})$, we have
$$
\frac{(\psi_{2n}\psi_{2m})^2(0,tq_\delta)} {\l_{2n}(tq_\delta)\!-\!\l_{2m}(tq_\delta)}=
\frac{(n\!+\!1)^{-\frac{1}{2}}(m\!+\!1)^{-\frac{1}{2}}+
O(n^{-\frac{1}{2}}m^{-1}+n^{-1}m^{-\frac{1}{2}})}{4\pi^2(n\!-\!m)}
$$
uniformly with respect to $t\!\in\![0,1]$\,. Note that the double sum $\sum_{n\ge 0 }
\sum_{m:m\ne n} \frac{O(\dots)}{4(n-m)}$ is absolutely convergent and so is equal to $0$.
Hence,
$$
\sum_{n\ge 0} (\mu_{2n}(q,b)\!-\!2\nh q_{2n}^+ - 2\pi^{-1/2}E_{n}b)
=\frac{b^2}{\pi^2}\lim_{k\to\infty}\sum_{n=0}^k\sum_{m=0,m\ne n}^{+\infty}
\frac{(n\!+\!1)^{-\frac{1}{2}}(m\!+\!1)^{-\frac{1}{2}}}{n-m}
$$
$$
= \frac{b^2}{\pi^2}\lim_{k\to\infty}\frac{1}{k^2}\sum_{n=0}^k\sum_{m=k+1}^{+\infty} \frac{1}
{\sqrt{\frac{n+1}{k}\cdot\frac{m+1}{k}}\left(\frac{n+1}{k}-\frac{m+1}{k}\right)} =
\frac{b^2}{\pi^2}\int_0^1dx\!\int_1^{+\infty}\!\!\!\frac{dy}{\sqrt{xy}\,(x\!-\!y)}
$$
$$
=\frac{4b^2}{\pi^2}\int_0^1dx\!\int_1^{+\infty}\!\!\!\frac{dy}{x^2\!-\!y^2} =
-\frac{2b^2}{\pi^2}\int_0^1\log\frac{1\!+\!x}{1\!-\!x}\,\frac{dx}{x}=
-\frac{4b^2}{\pi^2}\sum_{s=0}^{+\infty}\frac{1}{(2s\!+\!1)^2}= -\frac{b^2}{2}\,,
$$
since $\log\frac{1+x}{1-x}\!=\!2\sum_{s=0}^{+\infty}\frac{x^{2s+1}}{2s+1}$\,, $x\!\in\!(0,1]$
and $\sum_{s=0}^{+\infty}{(2s\!+\!1)^{-2}}\!=\! \frac{3}{4}\sum_{s=0}^{+\infty}{s^{-2}}\!=\!
\frac{3}{4}\cdot\frac{\pi^2}{6}\!=\!\frac{\pi^2}{8}$\,.
\end{proof}

\begin{proof}[{\bf Proof of Theorem \ref{*KasSD}.\ }]
Define  the meromorphic function
$$
f(\l)=\frac{\psi_+(0,\l,q)}{w_N(\l,q,b)}=
\frac{\psi_+(0,\l,q,b)}{\psi'_+(0,\l,q)\!-\!b\psi_+(0,\l,q)}\,,\qquad \l\!\in\!\C\,.
$$
By definition (\ref{SnDefN}) of $s_{2n}(q,b)$\,, we have
$$
\res_{\l=\l_{2n}(q,b)}f(\l)= \frac{\psi_+(0,\l_{2n}(q,b),q)}{\dot{w}_N(\l_{2n}(q,b),q,b)}=
\frac{(-1)^ne^{-s_{2n}(q,b)}}{\dot{w}_N(\l_{2n}(q,b),q,b)}\,\,.
$$
Note that
$$
f_0(\l)=\frac{\psi_+^0(0,\l)}{(\psi_+^0)'(0,\l)}=
\frac{1}{2}\,\cot\frac{(\l\!-\!1)\pi}{4}\cdot
\frac{\Gamma(\frac{1}{4}\,(\l\!+\!1))}{\Gamma(\frac{1}{4}\,(\l\!+\!3))}\,,\qquad
\res_{\l=\l_{2n}^0}f_0(\l)=2\pi^{-1/2}E_n\,.
$$

Put $|\l|\!=\!2k\!\to\!\infty$\,. Then, due to (\ref{EstimExact}), we have
$C_1k^{-\frac{1}{2}}\le|f_0(\l)|\le C_2k^{-\frac{1}{2}}$ for some absolute constants
$C_1\,,C_2\!>\!0$\,. Lemma \ref{*AEstim} gives
$$
\frac{\psi_+(0,\l,q)}{\psi_+^0(0,\l)}=
1+\frac{\psi_+^{(1)}(0,\l,q)}{{\psi_+^0(0,\l)}}+O(k^{-1})\,,\quad
\frac{\psi'_+(0,\l,q)}{(\psi_+^0)'(0,\l)}=
1+\frac{(\psi_+^{(1)})'(0,\l,q)}{(\psi_+^0)'(0,\l)}+O(k^{-1})
$$
Note that (\ref{TasJ}) implies
$\psi_+^{(1)}(0)\!=\!\int_0^{+\infty}(\psi_+^0\varphi^0)(t)q(t)dt$ and
$(\psi_+^{(1)})'(0)\!=\!-\int_0^{+\infty}(\psi_+^0\vartheta^0)(t)q(t)dt$\,, where we omit
$\lambda$ and $q$ for short. Since
$\psi_+^0(t)\!=\!\psi_+^0(0)\vartheta^0(t)\!+\!(\psi_+^0)'(0)\varphi^0(t)$\,, we have
$$
\frac{f(\lambda)}{f_0(\lambda)}=\frac{1+\frac{\psi_+^{(1)}}{\psi_+^0}\,(0)+O(k^{-1})}
{1+\frac{(\psi_+^{(1)})'}{(\psi_+^0)'}\,(0)- b f_0(\lambda)+O(k^{-1})} = 1+bf_0(\lambda)+
\frac{\int_0^{+\infty}(\psi_+^0)^2(t,\lambda)q(t)dt}{\psi_+^0(0)(\psi_+^0)'(0)}+O(k^{-1})\,.
$$
Therefore,
$$
f(\lambda)=f_0(\lambda)+b\cdot(f_0(\lambda))^2+h(\lambda,q)+O(k^{-\frac{3}{2}}),\quad {\rm
where} \quad
h(\lambda,q)=\frac{\int_0^{+\infty}(\psi_+^0)^2(t,\lambda)q(t)dt}{((\psi^0_+)')^2(0,\lambda)}\,.
$$
Applying the Cauchy Theorem in the disc $|\l|\!\le\!2k$ to this function and passing to the
limit as $k\!\to\!\infty$, we obtain
\begin{equation}
\label{xxRes} \sum_{n\ge 0}\biggl(\frac{(-1)^ne^{-s_{2n}(q,b)}}{\dot{w}_N(\l_{2n}(q,b),q,b)}-
2\pi^{-1/2}E_n\biggr)= b\cdot\sum_{n\ge 0}\res_{\l=\l_{2n}^0}(f_0(\lambda))^2+ \sum_{n\ge 0}
\res_{\l=\l_{2n}^0}h(\lambda)\,.
\end{equation}

Note that
$$
\res_{\l=\l_{2n}^0}(f_0(\l))^2=
-\frac{\ddot{g}_0(\lambda_{2n}^0)}{(\dot{g}_0(\lambda_{2n}^0))^3}\,, \quad {\rm where}\quad
g_0(\lambda)=\frac{1}{f_0(\lambda)}=\frac{(\psi_+^0)'}{\psi_+^0}\,(\lambda)\,,\ \
\lambda\!\in\!\C\,.
$$
On the other hand, the identity $g_0(\lambda_{2n}^0\!+\!\mu_{2n}(0,\beta))=\beta$ gives
$$
\frac{d\mu_{2n}(0,\beta)}{d\beta}\Big|_{\beta=0}=\frac{1}{\dot {g}_0
(\lambda_{2n}^0)}\,,\qquad \frac{d^2\mu_{2n}(0,\beta)}{d \beta^2}\Big|_{\beta=0}=
-\frac{\ddot{g}_0(\lambda_{2n}^0)}{(\dot {g}_0
(\lambda_{2n}^0))^3}=\res_{\l=\l_{2n}^0}(f_0(\l))^2\,.
$$
Due to Lemma \ref{*TraceNLN}, we have
\begin{equation}
\label{xxRes1} \sum_{n\ge 0} \res_{\l=\l_{2n}^0}(f_0(\l))^2= \sum_{n\ge 0}
\frac{d^2\mu_{2n}(0,\beta)}{d\beta^2}\Big|_{\beta=0}=-1\,.
\end{equation}

Furthermore,
$$
\sum_{n\ge 0} \res_{\l=\l_{2n}^0}h(\lambda)= \frac{1}{2\pi i}
\lim_{k\to\infty}\oint_{|\lambda|=2k}\!\!h(\lambda)d\lambda= \frac{1}{2\pi i}
\lim_{k\to\infty}\oint_{|\lambda|=2k}\!\!
\frac{\int_0^{+\infty}(\psi_+^0)^2(t,\lambda)q(t)dt}{((\psi^0_+)')^2(0,\lambda)}\,d\lambda\,.
$$
Using Lemma \ref{*AEstim}, asymptotics (\ref{EstimExact}) and the estimate
$\rho(t,\lambda)\!\ge\!\rho(t,|\lambda|)$, we obtain
$$
\biggl|\oint_{|\lambda|=2k}\!\!\!h(\lambda)d\lambda\biggr|\le
\int_0^{+\infty}\!\!|q(t)|R_k(t)dt\,,\qquad R_k(t)=\frac{C}{k^{\frac{1}{2}}\rho^2(t,2k)}\cdot
\oint_{|\lambda|={2k}}\!\!\!e^{-2\sigma(t,\lambda)}|d\lambda|\,,
$$
 where $C\!>\!0$ is some absolute constant. Since $\sigma(t,\lambda)\!\ge\!0$\,, we
have $R_k(t)=O(k^{\frac{1}{2}}\big/\rho^2(t,2k))$ uniformly with respect to $t$ and $k$.
Therefore, (\ref{BetaEstim}) yields the following estimate, which is uniform with respect to
$k$:
$$
\int_0^{+\infty}\!\!|q(t)|R_k(t)dt=
O(k^{1/2})\cdot\int_0^{+\infty}\frac{|q(t)|dt}{\rho^2(t,2k)}= O(k^{1/2})\cdot\beta(q,2k)=
O(\|q\|_{\bH_+})\,.
$$
On the other hand, due to definitions (\ref{ASRdef}), for any fixed $t\!>\!0$ we have
$$
R_k(t)=O(1)\cdot\int_{0}^{2\pi}\!e^{-2\sigma(t,2ke^{i\phi})}d\phi \to 0\,,\qquad
k\!\to\!\infty\,,
$$
since $e^{-2\sigma(t,2ke^{i\phi})}\!\le\!1$ and $\sigma(t,2ke^{i\phi})\!\to\!+\infty$ as
$k\!\to\!\infty$ for any $\phi\!\in\!(0,2\pi)$\,. Hence,
$$
\int_0^{+\infty}\!\!|q(t)|R_k(t)dt\to 0\,,\qquad k\!\to\!+\infty\,.
$$
This follows
\begin{equation}\label{xxRes2}
\sum_{n\ge 0} \res_{\l=\l_{2n}^0}h(\lambda)=0\,.
\end{equation}
Substituting (\ref{xxRes1}) and (\ref{xxRes2}) into (\ref{xxRes}), we obtain (\ref{KisEqual}).
\end{proof}

\begin{proof}[{\bf Proof of Theorem \ref{*H+NChar} (i).\ }]
Note that Theorem \ref{*AsymptN} gives $\widetilde{s}_{2n}(q)={\nc q_{2n}^+} +
\frac{1}{8}\,\pi E_n^2b^2 +\el2_{\frac{3}{4}+\delta}(n)$. Arguing in the same way as in the
proof of Theorem \ref{*H+DChar} (i), we obtain
$$
\sum_{n\ge 0}{\nc q_{2n}^+}z^n \equiv
P_+\biggl[\frac{(G_Nq)(\zeta)}{\sqrt{1\!-\!\overline{\zeta}}}\biggr]\equiv
-\frac{\pi}{2}P_+\biggl[\!\frac{(F_Dq)(\zeta)}{\sqrt{1\!-\!\overline{\zeta}}}\biggr]\equiv
-\frac{\pi}{2}P_+\biggl[\sqrt{-\zeta}\,\sum_{m\ge 0 }\nh q_{2m+1}^+\zeta^m\biggr]\,.
$$
Recall that $(\dlt Hq)(z)\!\equiv\!\sum_{m\ge 0}(\nh q_{2m}^+\!-\!\nh q_{2m+1}^+)z^m$\,. Then,
\begin{equation} \label{xxNCN}
\sum_{n\ge 0}{\nc q_{2n}^+}z^n \equiv -\frac{\pi}{2}P_+\biggl[\sqrt{-\zeta}\,\sum_{m\ge 0 }\nh
q_{2m}^+\zeta^m\biggr] + \frac{\pi}{2}P_+\biggl[\!\sqrt{-\zeta}\cdot(\dlt Hq)(\zeta)\biggr]\,,
\end{equation}
Due to Theorem \ref{*AsymptN}, we have
$$
\sum_{m\ge 0 }\nh q_{2m}^+z^m\equiv \frac{1}{2}\sum_{m\ge 0}\mu_{2m}(q,b)z^m+
\frac{b}{\pi^{1/2}}\sum_{m\ge 0}E_mz^m+\omega(z)\,,\quad \omega\in H^2_{\frac{3}{4}+\delta}\,.
$$
Note that Lemma \ref{*TraceNLN} yields $\omega(1)\!=\!\frac{1}{4}\,b^2$\,. Since $\sum_{m\ge
0}E_mz^m\equiv(1\!-\!z)^{-1/2}$, we have
$$ \sum_{m\ge 0 }\nh q_{2m}^+z^m\equiv \frac{1}{2}\sum_{m\ge 0}\mu_{2m}(q,b)z^m
+ \frac{\pi^{-1/2}b}{\sqrt{1\!-\!z}} + \frac{b^2}{4}+\omega_1(z)\,,\quad \omega_1\in\oo
H^2_{\frac{3}{4}+\delta}\,.
$$
Due to the identity $P_+[\,\sqrt{-\zeta}\big/\sqrt{1\!-\!\zeta}\,\,]\equiv
P_+[\,1\big/\sqrt{1\!-\!\overline{\zeta}}\,\,]\equiv 1$\,, we obtain
\begin{equation}
\label{xxNCN1} -\frac{\pi}{2}P_+\biggl[\!\sqrt{-\zeta}\,\sum_{m\ge 0 }\nh
q_{2m}^+\zeta^m\biggr]\equiv -\frac{\pi}{4}P_+\biggl[\!\sqrt{-\zeta}\,\sum_{m\ge 0
}\mu_{2m}(q,b)\zeta^m\biggr]+ \frac{\pi^{1/2}b}{2}-\frac{\pi
b^2}{8}P_+[\!\sqrt{-\zeta}\,]+\omega_2(z)\,,
\end{equation}
where $\omega_2(z)\equiv -\frac{\pi}{2}\,P_+[\sqrt{-\zeta}\omega_1(\zeta)]\in
H^2_{\frac{3}{4}+\delta}$\,.

We consider the second term in (\ref{xxNCN}). Recall that $(\dlt Hq)(1)\!=\!\frac{1}{4}\,q(0)$
and
$$
\frac{\pi}{2}\,P_+\biggl[\frac{(\dlt Hq)(\zeta)\!-\!(\dlt Hq)(1)}{\sqrt{-\zeta}}\biggr]\equiv
(\widetilde{H}q)(z)\equiv\sum_{n\ge 0}\nt q^+_n z^n\in H^2_{3/4}\,.
$$
Since $1\,\big/\sqrt{-\zeta}=\sqrt{1\!-\!\overline{\zeta}}\,\big/\sqrt{1\!-\!{\zeta}}$\,,
$\zeta\!\ne\!1$\,, we have
$$
(\dlt Hq)(z)\!-\!(\dlt Hq)(1)\equiv \frac{2}{\pi}\sqrt{1\!-\!z}\,
P_+\biggl[\frac{(\widetilde{H}q)(\zeta)}{\sqrt{1\!-\!\overline{\zeta}}}\biggr]\,.
$$
Using the identity $zP_+[f(\zeta)](z)\equiv P_+[\zeta f(\zeta)](z)- P_+[\zeta f(\zeta)](0)$\,,
we obtain
$$
\frac{\pi}{2}P_+\biggl[\!\sqrt{-\zeta}\cdot((\dlt Hq)(\zeta)\!-\!(\dlt Hq)(1))\biggr]\equiv
P_+\biggl[\frac{\,\ 1\!-\!\zeta}{\sqrt{1\!-\!\overline{\zeta}}}\,
P_+\biggl[\frac{(\widetilde{H}q)(\zeta)}{\sqrt{1\!-\!\overline{\zeta}}}\biggr]\biggr]
$$
$$
\equiv P_+\biggl[\frac{1}{\sqrt{1\!-\!\overline{\zeta}}}\biggl(P_+\biggl[
\frac{(1\!-\!\zeta)(\widetilde{H}q)(\zeta)}{\sqrt{1\!-\!\overline{\zeta}}}\biggr]+
P_+\biggl[\frac{\zeta(\widetilde{H}q)(\zeta)}{\sqrt{1\!-\!\overline{\zeta}}}\biggr](0)
\biggr)\biggr]
$$
\begin{equation}
\label{xxNCN2} \equiv
P_+\biggl[\frac{(1\!-\!\zeta)(\widetilde{H}q)(\zeta)}{1\!-\!\overline{\zeta}}\biggr] +
\sum_{k\ge 0}E_{k+1}\nt q_k^+ \equiv -\sum_{n\ge 1} \nt q_{n-1}^+z^n + \sum_{k\ge 0}E_{k+1}\nt
q_k^+\,.
\end{equation}

\pagebreak

Substituting (\ref{xxNCN1}) and (\ref{xxNCN2}) into (\ref{xxNCN}), we get
$$
\sum_{n\ge 0}\nc q_{2n}^+z^n \equiv -\frac{\pi}{4}\,P_+\biggl[\!\sqrt{-\zeta}\,\sum_{n\ge 0}
\mu_{2n}(q,b)\zeta^n\biggr] + \frac{\pi (q(0)\!-\!b^2)}{8}\,
P_+\left[\!\sqrt{-\zeta}\,\right]-\sum_{n\ge 0} \nt q_{n-1}^+z^n +\omega_2(z)\,,
$$
where
\begin{equation}
\label{NtQ-1Def}  \nt q_{-1}^+ =
-\sum_{k\ge 0}E_{k+1}\nt q_k^+ - \frac{\pi^{1/2}b}{2}\,.
\end{equation}
Using Theorem \ref{*AsymptN} and the identity
$\sqrt{-\zeta}=-\frac{2}{\pi}\sum_{l=-\infty}^{+\infty}\zeta^l\big/(2l\!-\!1)$\,, we obtain
$$
\widetilde{s}_{2n}(q)=\nc q_{2n}^+ +\frac{\pi E_n^2}{8}\,b^2 + \el2_{\frac{3}{4}+\delta}(n)=
\frac{1}{2}\sum_{m\ge 0} \frac{\mu_{2m}(q,b)}{2(n\!-\!m)\!-\!1}
-\frac{q(0)-2b^2}{4(2n\!-\!1)}+r_{2n}(q,b)\,,
$$
\begin{equation}
\label{RNasymp} r_{2n}(q,b)\!=\!-\nt q^+_{n-1} + \el2_{\frac{3}{4}+\delta}(n)\,.
\end{equation}
In particular, $\{r_{2n}(q,b)\}_{n=0}^\infty\!\in\!\el2_{3/4}$\,, since $\{\nt
q_{n-1}^+\}_{n=0}^\infty\!\in\!\el2_{3/4}$\,.
\end{proof}

\noindent {\bf Proof of Theorem \ref{*H+NChar} (ii).} As in the proof of Theorem
\ref{*H+DChar}, we prove that the mapping $\Phi$ is 1) injective; 2) analytic; 3) local
isomorphism and 4) surjective, where
$$
\Phi: (q;b)\mapsto \left(\{\mu_{2n}(q,b)\}_{n=0}^\infty\,;q(0)-2b^2\,;
\{r_{2n}(q,b)\}_{n=0}^\infty\right)\,,\qquad
\bH_+\!\times\R\to\cS_N\times\R\times\el2_{3/4}\,.
$$

\begin{proof}[{\bf 1. Uniqueness Theorem.}] Suppose that
$$
\left(\{\mu_{2n}(p,a)\}_{n=0}^\infty\,,p(0)\!-\!2a^2\,, \{r_{2n}(p,a)\}_{n=0}^\infty\right)=
\left(\{\mu_{2n}(q,b)\}_{n=0}^\infty\,,q(0)\!-\!2b^2\,, \{r_{2n}(q,b)\}_{n=0}^\infty\right)
$$
for some potentials $p,q\!\in\!\bH_+$ and some real constants $a,b\!\in\!\R$\,. It follows
$$
\l_{2n}(p)\!=\!\l_{2n}(q)\,,\quad s_{2n}(p)\!=\!s_{2n}(q)\,,\quad n\!\ge\!0\,.
$$
Theorem \ref{*KasSD} yields $a\!=\!b$\,. The rest of the proof repeats the proof of Theorem
\ref{*H+DChar} with the function $\vartheta(x,\lambda,q)\!-\!b\varphi(x,\lambda,q)$ instead of
$\varphi(x,\lambda,q)$.
\end{proof}

\begin{proof}[{\bf 2. Analyticity of $\Phi$\,.}]
Arguments, similar to the proof of (\ref{TildeQIso}), follow that the mappings
$$
\begin{array}{c}\displaystyle
q\mapsto (F_Nq\,;\dlt H q) \mapsto \left(\{\nh q_{2n}^+\}_{n=0}^\infty\,;q(0)\,;\{\nt
q_n^+\}_{n=0}^\infty\right)\,,\cr \displaystyle \bH_+\ \leftrightarrow\ \po H^2_{3/4}\times
\po H^2_{3/4} \quad \leftrightarrow \quad\ \ \cH\times\R\times\el2_{3/4}\,,\qquad\ \
\end{array}
$$
are linear isomorphisms. Also, it is clear that the mapping
$$
\left(\{\nh q_{2n}^+\}_{n=0}^\infty\,;b\right)\mapsto \left(\{2\nh q_{2n}^+ \!+\!
2\pi^{-\frac{1}{2}}E_nb\}_{n=0}^\infty\,;b\right)
$$
is a linear isomorphism between $\cH\times \R$ and $\cH\times\R$\,. Furthermore, since $\nt
q_{-1}^+$ is a linear combination of $\nt q_n^+$\,, $n\!\ge\!0$\,, and $b$ (see
(\ref{NtQ-1Def})), the mapping
$$
\left(b\,;\{\nt q_n^+\}_{n=0}^{\infty}\right)\mapsto \{\nt q_{n-1}^+\}_{n=0}^{+\infty}
$$
is a linear isomorphism between $\R\times\el2_{3/4}$ and $\el2_{3/4}$\,. Combining these
isomorphisms, we obtain that the linear operator
$$
\Phi_0:(q;b)\mapsto\left(\{2\nh q_{2n}^+ \!+\! 2\pi^{-\frac{1}{2}}E_nb\}_{n=0}^\infty\,;
q(0)\,; \{\nt q_{n-1}^+\}_{n=0}^\infty \right)\,,\quad
\Phi_0:\bH_+\!\times\R\leftrightarrow\cH\times\R\times\el2_{3/4}
$$
and its inverse are bounded. Recall that Theorem \ref{*AsymptN} gives
$$
\mu_{2n}(q,b)=2\nh q_{2n}^+ + 2\pi^{-1/2}E_nb+ \el2_{\frac{3}{4}+\delta}(n)\,, \qquad
r_{2n}(q)=-\nt q_{n-1}^+ + \el2_{\frac{3}{4}+\delta}(n)\,.
$$
In other words, if we put $\Phi\!=\!\Phi_0\!+\!\Phi_1$\,, then $\Phi_1$ maps $\bH_+\!\times\R$
into the space $\el2_{\frac{3}{4}+\delta}\times\R\times\el2_{\frac{3}{4}+\delta}$\,. Moreover,
$\Phi_1$ is locally bounded. Using Lemma \ref{*AAnalyticity} by the same way as in Sect. 4, we
obtain that $\Phi_1$ is analytic and hence $\Phi$ is analytic too.
\end{proof}

\begin{proof}[{\bf 3. Local Isomorphism.}] The proof is similar to the case of
Dirichlet boundary condition in Sect. 4. Due to Lemma \ref{*AAnalyticity}, the following
identities are fulfilled:
\begin{equation}
\label{GradIdentN} \frac{\partial\mu_{2n}(q,b)}{\partial q(t)}= \psi_{2n}^2(t,q,b)\,,\qquad
\frac{\partial s_{2n}(q)}{\partial q(t)}= (\psi_{2n}\chi_{2n})(t,q,b)\,,
\end{equation}
where $\psi_{2n}(\cdot,q,b)$ is $n$-th normalized eigenfunction of $T_b$ and
$\chi_{2n}(\cdot,q,b)$ is some special solution of Eq. \nolinebreak (\ref{PertEq}) for
$\l\!=\!\l_{2n}(q,b)$ such that $\{\chi_{2n+1}\,,\psi_{2n+1}\}\!=\!1$\,. In particular,
$(\psi_{2n}\chi_{2n})'\!\in\!\bH_+$\,. Also,
\begin{equation}
\label{GradIdentNb} \frac{\partial\mu_{2n}(q,b)}{\partial b} = \psi_{2n}^2(0,q,b)\,,\qquad
\frac{\partial s_{2n}(q)}{\partial b} = (\psi_{2n}\chi_{2n})(0,q,b)\,.
\end{equation}

Note that for each $q\!\in\!\bH_+$\,, $b\!\in\!\R$, the Fr\'echet derivative
$$
d_{(q;b)}\Phi=\Phi_0+d_{(q;b)}\Phi_1:\bH_+\!\times\R\to\cH\times\R\times\el2_{3/4}
$$
is a Fredholm operator since $\Phi_0^{-1}$ is bounded and $d_{(q;b)}\Phi_1$ maps
$\bH_+\!\times\R$ into $\el2_{\frac{3}{4}+\delta}\times\R\times\el2_{\frac{3}{4}+\delta}$ and
so is compact as an operator from $\bH_+$ to
$\el2_{\frac{3}{4}}\times\R\times\el2_{\frac{3}{4}}$\,. Hence, it is sufficient to show that
\begin{equation}
\label{ClImPhi} \cH\times\R\times\el2_{3/4}=\overline{\Ran d_{(q;b)}\Psi}\,.
\end{equation}

Using (\ref{GradIdentN}), (\ref{GradIdentNb}) and Lemma \ref{*AGradProd} (ii), we get
$$
\biggl(2(\psi_{2n}^2)'(q,b),\frac{\partial\mu_{2n}(q,b)}{\partial q}\biggr)_{+} +\
\psi_{2n}^2(0,q,b)\frac{\partial\mu_{2n}(q,b)}{\partial b}\,=\,0\,,
$$
$$
\biggl(2(\psi_{2n}^2)'(q,b),\frac{\partial s_{2n}(q,b)}{\partial q}\biggr)_{+} +\
\psi_{2n}^2(0,q,b)\frac{\partial s_{2n}(q,b)}{\partial b}\,=\,\delta_{nm}\,,
$$
Moreover, note that
$$
(d_{(q;b)}(q(0)\!-\!2b^2))(2(\psi_{2n}^2)'(q,b);\psi_{2n}^2(0,q,b))=
2(\psi_{2n}^2)'(0,q,b)\!-\!4b\psi_{2n}^2(0,q,b)= 0\,,
$$
since $\psi'_{2n}(0,q,b))\!=\!b\psi_{2n}(0,q,b)$\,. Hence, definition (\ref{RnDefN}) gives
$$
(d_{(q;b)}\Phi)\left(2(\psi_{2n}^2)'(\cdot,q,b)\,;\psi_{2n}^2(0,q,b)\right) = \left({\bf 0}\,;
0\,;{\bf e}_n\right)\,.
$$
Therefore,
$$
\left\{({\bf 0};0)\right\}\times\el2_{3/4} \subset\overline{\Ran d_{(q;b)}\Phi}\,.
$$
We consider the lowest eigenvalue $\l_1(q)$ of the operator $T_D$ (with the same potential
$q$) and the function
$$
\xi(t)=((\vartheta\!-\!b\varphi)\varphi)'(t,\l_1(q),q)\,.
$$
Note that $\xi(0)\!=\!1$\,, $\xi\!\in\!\bH_+$ and
$\left(\xi,\psi_{2m}^2(q)\right)_{+}\!=\!0$\,. Hence,
$$
(d_{(q;b)}\Phi)(\xi\,;0) = \left({\bf 0}\,; 1\,;(d_{(q;b)}r)(\xi\,;0)\right)\,.
$$
This implies
$$
\{{\bf 0}\}\times\R\times\el2_{3/4}\subset\overline{\Ran d_{(q;b)}\Phi}\,.
$$
Furthermore, using Lemma \ref{*AGradProd} (ii) again, we obtain
$$
(d_{(q;b)}\Phi)\left(-2(\psi_{2m}\chi_{2m})'\,;-(\psi_{2m}\chi_{2m})(0)\right) = \left({\bf
e}_m\,; 2\,; (d_{(q;b)}r)(-2(\psi_{2m}\chi_{2m})'\,;-(\psi_{2m}\chi_{2m})(0))\right)\,.
$$
Since the set of finite sequences is dense in $\cH_0$\,, it yields
$$
\cH_0\times\R\times\el2_{3/4}\subset\overline{\Ran d_{(q,b)}\Phi}\,.
$$
In conclusion, we consider an arbitrary function $\zeta\!\in\!\bH_+$ such that
$\int_{\R_+}\zeta(t)dt\!\ne\!0$\,. Due to asymptotics (\ref{IntCH0AsymptN}), we have
$$
(d_{(q,b)}\Phi)(\zeta\,;0)\notin\cH_0\times\R\times\el2_{3/4}\,.
$$
Hence, (\ref{ClImPhi}) holds true since the codimension of $\cH_0$ in $\cH$ is equal to $1$.
\end{proof}

\begin{proof}[{\bf 4. $\Phi$ is surjective.}] The following Lemma is an analogue of Lemma
\ref{*S2k+1Change}:
\begin{lemma}
\label{*S2kChange} Let $q\!\in\!\bH_+$\,, $b\!\in\!\R$\,, $n\!\ge\!0$ and $t\!\in\!\R$\,.
Denote
$$
q_{2n}^t(x)=q(x)-2\frac{d^2}{dx^2}\log\eta_{2n}^t(x,q,b)\,,\qquad
b_{2n}^{\,t}=b+(1\!-\!e^{-t})\,\psi_{2n}^2(0,q,b)\,,
$$
where
$$
\eta_{2n}^t(x,q,b)= 1+(e^t\!-\!1)\int_x^{+\infty}\!\!\psi_{2n}^2(s,q,b)ds\,.
$$
Then $q_{2n}^t\!\in\!\bH$ and
$$
\l_{2m}(q_{2n}^t\,,b_{2n}^{\,t})=\l_{2m}(q,b)\,,\qquad
s_{2m}(q_{2n}^t\,,b_{2n}^{\,t})=s_{2m}(q,b)+t\delta_{nm}
$$
for all $m\!\ge\!0$\,. Moreover, $q_{2n}^t(0)\!-\!2(b_{2n}^{\,t})^2\!=\!q(0)\!-\!2b^2$.
\remark {\rm In this case definition (\ref{RnDefN}) yields
$r_{2m}(q_{2n}^t\,,b_{2n}^{\,t})\!=\!r_{2m}(q,b)\!+\!t\delta_{nm}$\,.}
\end{lemma}
\begin{proof}
The crucial point is the changing of $b$\,, i.e. the changing of the boundary condition. It is
known (see \cite{PT}, \cite{MT}, \cite{CKK}) that for each $m\!\ge\!0$ the function
$$
\widetilde{\psi}_{2m}(x)=
\psi_{2n}(x,q,b)-(e^t\!-\!1)\frac{\psi_{2n}(x,q,b)}{\eta_{2n}^t(x,q,b)}
\int_x^{+\infty}\psi_{2n}(s,q,b)\psi_{2m}(s,q,b)ds
$$
is a solution of the equation $-\psi''+x^2\psi+q_{2n}^t(x)\psi=\l_{2m}(q,b)\psi$\,. Direct
calculations show that the identities
$\widetilde{\psi}'_{2m}(0)\!=\!b_{2n}^t\widetilde{\psi}_{2m}(x)$ holds true for all
$m\!\ge\!0$\,. Hence, $\l_{2m}(q_{2n}^t\,,b_{2n}^{\,t})=\l_{2m}(q,b)$\,. The other equalities
are similar to Lemma \ref{*S2k+1Change}.
\end{proof}

The rest of the proof is the same as in Sect. 4. We consider an arbitrary spectral data
$(h^*;u^*;c^*)\in\cS_D\times\R\times\el2_{3/4}$\,. Due to Theorem \ref{*H+2SpecChar}, there
exists potential $q^*\!\in\!\bH_+$ such that
$$
\mu_{2n}(q^*)=\mu_{2n}(q^*,0)=h^*_n\,,\ \ n\!\ge\!0\,,\qquad q^*(0)=u^*\,.
$$
Since $\Phi$ is a local isomorphism and the set of finite sequences is dense in $\cH_0$\,,
there exists a finite number of values $t_0\,,t_2\,,\dots,t_{2k}$ such that
$$
\left(h^*;u^*;(c^*_0\!-\!t_0\,,\dots,c^*_{2k}\!-\!t_{2k}\,,
c^*_{2k+2}\,,c^*_{2k+4}\,,\dots)\right)= \Phi(q_{k+1}\,;b_{k+1}) \in \Phi(\bH_+\!\times\R)\,.
$$
Using Lemma \ref{*S2kChange} $k\!+\!1$ times step by step, we obtain potential
$q_{0}\!\in\!\bH_+$ and constant $b_{0}\!\in\!\R$\,, such that
$\Psi(q_0\,;b_0)\!=\!(h^*;u^*;c^*)$. The proof is finished.
\end{proof}

\renewcommand{\thesection}{A}
\section{Appendix}
\setcounter{equation}{0}
\renewcommand{\theequation}{A.\arabic{equation}}

Here we collect some technical results from \cite{CKK}, \cite{CKK1} which are essentially used
above.

\noindent {\bf A.1 The unperturbed equation. } It is well-known that for each $\l\!\in\!\C$
the equation
\begin{equation}
\label{HosEq} -\p''+x^2\p=\l\p\,.
\end{equation}
has the solution $\p_{+}^0(x,\l)=D_{\frac{\l-1}{2}}(\sqrt{2}x)$\,, where $D_{\mu}(x)$ is the
Weber function (or the parabolic cylinder function, see \cite{B}). Moreover, for each $x$ the
functions $\p_{+}^0(x,\cdot)$ and $(\p_{+}^0)'(x,\cdot)$ are entire and the following
asymptotics are fulfilled:
\begin{equation}
\label{Psi+0Asympt}
\begin{array}{c}\displaystyle
\p_{+}^0(x,\l)=(\sqrt{2}x)^{\frac{\l-1}{2}}e^{-\frac{x^2}{2}} \left(1+O(x^{-2})\right)\,,\quad
x\to +\infty\,, \vphantom{\frac{1}{\sqrt{2}}}\cr \displaystyle \label{Psi+0DerAsympt}
(\p_{+}^0)'(x,\l)=-\frac{1}{\sqrt{2}}(\sqrt{2}x)^{\frac{\l+1}{2}}e^{-\frac{x^2}{2}}
\left(1+O(x^{-2})\right)\,,\quad x\to +\infty\,,
\end{array}
\end{equation}
uniformly with respect to $\l$ on bounded domains. Note that (see \cite{B})
\begin{equation}
\label{Psi+00lambda}
\begin{array}{c}\displaystyle\psi_+^0(0,\l)=
D_{\frac{\l-1}{2}}(0)=2^{\frac{\l-1}{4}}{\Gamma(\frac{1}{2})\/\Gamma(\frac{3-\l}{4})}=
\cos\frac{(\l\!-\!1)\pi}{4}\,\cdot\,
\frac{2^\frac{\l-1}{4}}{\sqrt{\pi}}\,\,\Gamma\!\biggl(\frac{\l\!+\!1}{4}\biggr)\,,\cr
\displaystyle (\psi_+^0)'(0,\l)=\sqrt{2}D'_{\frac{\l-1}{2}}(0)=
2^{\frac{\l-1}{4}}{\Gamma(-\frac{1}{2})\/\Gamma(\frac{1-\l}{4})}=
\sin\frac{(\l\!-\!1)\pi}{4}\,\cdot\,
\frac{2^\frac{\l+3}{4}}{\sqrt{\pi}}\,\,\Gamma\!\biggl(\frac{\l\!+\!3}{4}\biggr)\,.
\end{array}
\end{equation}
Let $J^0(x,t;\l)$ be the solution of (\ref{HosEq}) such that $J^0(t,t;\l)= 0$,
$(J^0)'_x(t,t;\l)= 1$\,. Note that
\begin{equation} \label{TasJ}
J^0(0,t;\l)= -\varphi^0(t,\l)\,,\qquad (J^0)'_x(0,t;\l)= \vt^0(t,\l)\,.
\end{equation}
In order to estimate $\psi_+^0$ and $J^0$\,, we introduce real-valued functions
$$
\r(x,\l)= 1+|\l|^{1/12}+|x^2-\l|^{1/4}\,,
$$
\begin{equation}
\label{ASRdef} a(\l)=
\biggl|\frac{\l}{2e}\biggr|^{\frac{\Re\l}{4}}\!e^{\frac{\pi-\phi}{4}\Im\l}, \quad\quad
\l=|\l|e^{i\phi},\ \ \phi\in[0,2\pi)\,,
\end{equation}
$$
\s(x,\l)= \Re\int_0^x\!\!\sqrt{y^2-\l}\,dy\,,\quad x\!\ge\!0\,,
$$
where $\sqrt{y^2\!-\!\l}=y\!+\!o(1)$ as $y\!\to\!+\infty$ (note that it follows
$\Re\sqrt{y^2\!-\!\l}\!\ge\!0$\,, if $y\!\ge\!0$).

\begin{lemma} \label{*AUnpertEstim}
For all $(x,t,\l)\in\R_+\times\R_+\times\C$ the following estimates are fulfilled:
\begin{equation}\label{Psi0Estim}
|\p_{+}^0(x,\l)|\le C_0 a(\l)\cdot \frac{e^{-\s(x,\l)}}{\r(x,\l)}\,, \qquad\qquad\
|(\p_{+}^0)'(x,\l)|\le C_0 a(\l)\cdot \r(x,\l)e^{-\s(x,\l)}\,,
\end{equation}
$$
|J^0(x,t;\l)|\le \frac{C_1}{\r(x,\l)\r(t,\l)}\, e^{|\s(x,\l)-\s(t,\l)|}\,, \quad
|(J^0)'_x(x,t;\l)|\le C_1\frac{\r(x,\l)}{\r(t,\l)}\, e^{|\s(x,\l)-\s(t,\l)|}\,,
$$
where $C_0,\ C_1$ are some absolute constants.
\end{lemma}
\begin{proof} See Lemmas 2.1 and 2.3 \cite{CKK1}. Note that the proof is based on
the result of \cite{O}. \end{proof}

\remark If $x\!=\!0$ and $|\lambda|\!\ge\!1$, then\begin{footnote}{Here and below $f\asymp g$
means that $C_1|f|\le|g|\le C_2|f|$ for some absolute constants $C_1,C_2>0$.}\end{footnote}
$\s(0,\lambda)\!=\!0$ and $\rho(0,\lambda)\!\asymp |\lambda|^{1/4}$. Identities
(\ref{Psi+00lambda}) and routine calculations follow
\begin{equation}\label{EstimExact}
\begin{array}{ll}\displaystyle
|\psi_+^0(0,\lambda)|\asymp |\lambda|^{-{1}/{4}}a(\lambda)\,, & {\rm if}\ \ |\lambda|=k\ne
4n\!+\!3\,,\vphantom{\big|_{\big|}}\cr\displaystyle |(\psi_+^0)'(0,\lambda)|\asymp
|\lambda|^{{1}/{4}}a(\lambda)\,,& {\rm if}\ \ |\lambda|=k\ne 4n\!+\!1\,,
\vphantom{\big|^{\big|}}\end{array}\ \ k,n\!\in\!\N\,.
\end{equation}
In other words, the estimates (\ref{Psi0Estim}) of $|\psi_+(0,\lambda)|$ and
$|\psi'_+(0,\lambda)|$ are exact on these contours.

\vskip 6pt

\noindent {\bf A.2 The perturbed equation. } The solutions $\psi_+\,,\vartheta\,,\varphi$ of
the perturbed equation
\begin{equation}
\label{AhoEq} -\p''+x^2\p+q(x)\p=\l\p,\quad \l\in \C\,,
\end{equation}
can be constructed by iterations:
\begin{equation}
\label{Psi+asSum} \p_{+}(x,\l,q)={\mathop{\sum}\limits_{n\ge 0}} \p_{+}^{(n)}(x,\l,q)\,, \quad
\p_{+}^{(n+1)}(x,\l,q)= -\int_x^{+\infty}\!\!J^0(x,t;\l)\p_{+}^{(n)}(t,\l,q)q(t)dt\,,
\end{equation}
\begin{equation}
\label{T12asSum} \vt_{1,2}(x,\l,q)={\mathop{\sum}\limits_{n\ge
0}}\vt_{1,2}^{(n)}(x,\l,q),\qquad \vt_{1,2}^{(n+1)}(x,\l,q)=
\int_0^x\!J^0(x,t;\l)\vt_{1,2}^{(n)}(t,\l,q)q(t)dt\,,
\end{equation}
where we use the notations $\vt_1\!=\!\vt$ and $\vt_2\!=\!\varphi$ for short.

\pagebreak

Introduce functions
$$
\beta_{+}(x,\l,q)= C_1\int_x^{+\infty}\!\frac{|q(t)|dt}{\r^2(t,\l)}\,, \qquad
\beta_{0}(x,\l,q)= C_1\int_0^x\frac{|q(t)|dt}{\r^2(t,\l)}dt\,.
$$
It is easy to see (\cite{CKK} Lemma \nolinebreak 5.5) that
\begin{equation}
\label{BetaEstim} \beta(\lambda,q)=\beta_{+}(x,\l,q)\!+\!\beta_{0}(x,\l,q)=
C_1\int_0^{+\infty}\frac{|q(t)|dt}{\r^2(t,\l)}=O(|\lambda|^{-1/2}\|q\|_{\bH_+})\,.
\end{equation}
\begin{lemma} \label{*AEstim}
For all $(x,\l,q)\in\R_+\!\times\C\times\bH_{+\C}$ the following estimates are fulfilled:
$$
|\p_{+}^{(n)}(x,\l,q)|\le C_0 a(\l)\frac{e^{-\s(x,\l)}}{\r(x,\l)}\cdot
\frac{\beta_{+}^n(x,\l,q)}{n!}\,,
$$
$$
|\vt_j^{(n)}(x,\l,q)|\le \frac{2 C_1}{(1\!+\!|\l|^{1/4})^{2j-3}}
\cdot\frac{e^{\s(x,\l)}}{\r(x,\l)}\cdot \frac{\beta_0^n(x,\l,q)}{n!}\,,\quad j\!=\!1,2\,.
$$
In particular, series (\ref{Psi+asSum}), (\ref{T12asSum}) converge uniformly on bounded
subsets of $\R_+\times\C\times\bH_{+\C}$\,. Moreover, the similar estimates with $\rho(x,\l)$
instead of $\frac{1}{\rho(x,\l)}$ in right-hand sides hold true for the values
$|(\p_{\pm}^{(n)})'(x,\l,q)|$ and $|(\vt_j^{(n)})'(x,\l,q)|$\,.
\end{lemma}
\begin{proof} See \cite{CKK1} Lemma 3.1 and \cite{CKK} Lemmas 5.2, 5.3.\end{proof}

\begin{corollary}\label{*ADotEstim}
For all $(\l,q)\in\C\times\bH_{+\C}$\,, $n,m\!\ge\!0$ and some absolute constant $C\!>\!0$ the
following estimates are fulfilled:
\begin{equation}\label{APsiNMEstim}
\begin{array}{c}\displaystyle
\biggl|\frac{\partial^m\p_{+}^{(n)}(0,\l,q)}{\partial\lambda^m}\biggr| \le
\frac{m!C^{n+m+1}\|q\|_{\bH_+}^n}{n!}\cdot \frac{\log^m(|\lambda|\!+\!2)\cdot
a(\l)}{(|\lambda|+1)^{\frac{n}{2}+\frac{1}{4}}}\,, \vphantom{\Big|_{\Big|}}\cr\displaystyle
\biggl|\frac{\partial^m(\p_{+}^{(n)})'(0,\l,q)}{\partial\lambda^m}\biggr| \le
\frac{m!C^{n+m+1}\|q\|_{\bH_+}^n}{n!}\cdot \frac{\log^m(|\lambda|\!+\!2)\cdot
a(\l)}{(|\lambda|+1)^{\frac{n}{2}-\frac{1}{4}}}\,.\vphantom{\Big|^{\Big|}}
\end{array}
\end{equation}
\end{corollary}
\begin{proof} Note that $\sigma(0,\lambda)\!=\!0$ and
$\rho(0,\lambda)\!\asymp\!1\!+\!|\lambda|^{1/4}$\,. Hence, Lemma \ref{*AEstim} and
(\ref{BetaEstim}) give (\ref{APsiNMEstim}) for $m\!=\!0$\,. Recall that
$\psi_+^{(n)}(0,\lambda,q)$, $(\psi_+^{(n)})'(0,\lambda,q)$ are entire functions. Therefore,
the integration over the contour
$\lambda(\phi)\!=\!\lambda+e^{i\phi}\log^{-1}(|\lambda|\!+\!2)$\,, $\phi\!\in\![0,2\pi]$\,,
together with the simple estimate $a(\lambda(\phi))\!=\!O(a(\lambda))$ follows
(\ref{APsiNMEstim}) in the general case $m\!>\!0$\,.
\end{proof}

Let
$$
\widetilde{\beta}_{+}(x,q)=\frac{1}{x^2\!+\!1}\,+
\int_x^{+\infty}\biggl|\frac{q(t)}{t}\biggr|\,dt\,.
$$
The following asymptotics as $x\!\to\!+\infty$ are fulfilled uniformly on bounded subsets of
$\C\times\bH_{+\C}$ (see \cite{CKK} p.139 and p.169):
\begin{equation}
\label{Psi+Asympt1}
\begin{array}{c}\displaystyle
\p_{+}(x,\l,q)=(\sqrt{2}x)^{\frac{\l-1}{2}}e^{-\frac{x^2}{2}}
(1\!+\!O(\widetilde{\beta}_+(x,q)))\,, \vphantom{\frac{1}{\sqrt{2}}}\cr \displaystyle
\p'_+(x,\l,q)=-\frac{1}{\sqrt{2}}(\sqrt{2}x)^{\frac{\l+1}{2}}e^{-\frac{x^2}{2}}
(1\!+\!O(\widetilde{\beta}_+(x,q)))\,.
\end{array}
\end{equation}
Moreover, if $\chi_{+}(x,\l,q)$ is an arbitrary solution of (\ref{AhoEq}) such that
$k=\{\chi_{+},\p_{+}\}\ne 0$\,, then
\begin{equation}
\label{Chi+Asympt1}
\begin{array}{c}\displaystyle
\chi_{+}(x,\l,q)=-\frac{k}{\sqrt{2}} (\sqrt{2}x)^{\frac{-\l-1}{2}}e^{\frac{x^2}{2}}
(1\!+\!O(\widetilde{\beta}_{+}(x,q)))\,, \cr
\displaystyle\vphantom{\frac{1^|}{\sqrt{2}}}\chi'_{+}(x,\l,q)\!=\!-\frac{k}{2}
(\sqrt{2}x)^{\frac{-\l+1}{2}}e^{\frac{x^2}{2}} (1\!+\!O(\widetilde{\beta}_{+}(x,q)))\,.
\end{array}
\end{equation}
\remark If $q\!\in\!\bH_{+\C}$\,, then (\ref{Psi+Asympt1}), (\ref{Chi+Asympt1}) give
$(\psi_+\chi_+)'\!\in\!\bH_{+\C}$ (see \cite{CKK} p. 172).

\vskip 6pt  \noindent{\bf A.3 Analyticity of spectral data and its gradients.}
\begin{lemma}\label{*AAnalyticity}
For each $(q;b)\!\in\!\bH_+\!\times\!\R$ all spectral data $\lambda_{2n+1}(q)$, $s_{2n+1}(q)$,
$\lambda_{2n}(q,b)$, $s_{2n}(q,b)$ extend analytically to some complex ball
$\{(p;a)\!\in\!\bH_{+\C}\times\C:\|p-q\|_{\bH_{+\C}}^2\!+|a-b|^2\!<\!R^2(q,b)\}$. Moreover,
its gradients are given by
$$
\frac{\partial\lambda_{2n+1}(q)}{\partial q(t)}= \psi_{2n+1}^2(t,q)\,,\qquad\quad
\frac{\partial\lambda_{2n}(q,b)}{\partial q(t)}= \psi_{2n}^2(t,q,b)\,,
$$
$$
\frac{\partial s_{2n+1}(q)}{\partial q(t)}= (\psi_{2n+1}\chi_{2n+1})(t,q)\,, \qquad
\frac{\partial s_{2n}(q,b)}{\partial q(t)}= (\psi_{2n}\chi_{2n})(t,q,b)\,,
$$
where $\psi_{2n+1}$ is the $n$-th normalized eigenfunction of $T_D$\,, $\psi_{2n}$ is the
$n$-th normalized eigenfunction of $T_b$, and
$$
\chi_{2n+1}(t,q)= \frac{\vartheta(t,\l_{2n+1}(q),q)}{\psi'_{2n+1}(0,q)}-
\frac{\dot{\psi}'_+}{\psi'_+}\,(0,\l_{2n+1}(q),q)\cdot\psi_{2n+1}(t,q)\,,
$$
$$
\chi_{2n}(t,q,b)= -\frac{\varphi(t,\l_{2n}(q,b),q)}{\psi_{2n}(0,q,b)}-
\frac{\dot{\psi}_+}{\psi_+}\,(0,\l_{2n}(q,b),q)\cdot \psi_{2n}(t,q,b)\,.
$$
Furthermore,
$$
\frac{\partial\lambda_{2n}(q,b)}{\partial b}= \psi_{2n}^2(0,q,b)\,,\qquad \frac{\partial
s_{2n}(q,b)}{\partial b}= (\psi_{2n}\chi_{2n})(0,q,b)\,.
$$
\end{lemma}
\begin{proof} The proof of the analyticity repeats the proof of \cite{CKK} Lemma 2.3 (p. 172). In
order to calculate gradients note that standard arguments (see \cite{CKK} Lemma 5.6) give
$$
\frac{\partial \psi_+(0,\lambda,q)}{\partial q(t)}=(\vp\p_+)(t,\lambda,q),\qquad
\frac{\partial \psi'_+(0,\lambda,q)}{\partial q(t)}=-(\vt\p_+)(t,\lambda,q),\quad t\!\ge\! 0.
$$
Applying the implicit function Theorem to the equation $\psi_+(0,\lambda_{2n+1}(q),q)\!=\!0$
and using the identity
$\int_{\R_+}{\psi_+^2}(t,\lambda,q)dt\!=\!\{\psi_+,\dot{\psi}_+\}(0,\lambda,q)$, we obtain
$$
\frac{\partial \lambda_{2n+1}(q)}{\partial q(t)}=-\frac{\partial \psi_+(0)/\partial
q(t)}{\partial \psi_+(0)/\partial \lambda}= -\frac{(\varphi\psi_+)(t)}{\dot{\psi}_+(0)}=
-\frac{\psi_+^2(t)}{\psi_+'(0)\dot{\psi}_+(0)}=
\frac{\psi_+^2(t)}{\{\psi_+,\dot{\psi}_+\}(0)}=\psi_{2n+1}^2(t)
$$
(here and below we omit $\lambda\!=\!\lambda_{2n+1}(q)$ and $q$ for short). Moreover,
$$
\frac{\partial s_{2n+1}(q)}{\partial q(t)} =
-\frac{\partial\log|\psi'_+(0,\lambda_{2n+1}(q),q)|}{\partial q(t)}=
\frac{(\vartheta\psi_+)(t)-\dot{\psi}'_+(0)\cdot\psi_{2n+1}^2(t)}{\psi'_+(0)}=
(\psi_{2n+1}\chi_{2n+1})(t)\,.
$$

By the same way, the identity $(\psi'_+\!-b\psi_+)(0,\lambda_{2n}(q,b),q)$ follows
$$
\frac{\partial \lambda_{2n}(q)}{\partial q(t)}=-\frac{\partial (\psi'_+\!-b\psi_+)(0)/\partial
q(t)}{\partial (\psi'_+\!-b\psi_+)(0)/\partial \lambda}=
\frac{(\vartheta+b\varphi)(t)\psi_+(t)}{\dot{\psi}'_+(0)-b\dot{\psi}_+(0)}=
\frac{\psi_+^2(t)}{\psi_+(0)(\dot{\psi}'_+(0)-b\dot{\psi}_+(0))}=\psi_{2n}^2(t)\,,
$$
$$
\frac{\partial \lambda_{2n}(q)}{\partial b}=
\frac{\psi_+(0)}{\dot{\psi}'_+(0)-b\dot{\psi}_+(0)}=
\frac{\psi_+^2(0)}{\{\psi_+,\dot{\psi}_+\}(0)}=\psi_{2n}^2(0)\,,
$$
Furthermore,
$$
\frac{\partial s_{2n}(q)}{\partial q(t)} =
-\frac{\partial\log|\psi_+(0,\lambda_{2n}(q),q)|}{\partial q(t)}=
-\frac{(\varphi\psi_+)(t)+\dot{\psi}_+(0)\cdot\psi_{2n}^2(t)}{\psi_+(0)}=
(\psi_{2n}\chi_{2n})(t)\,,
$$
$$
\frac{\partial s_{2n}(q)}{\partial b} =
-\frac{\partial\log|\psi_+(0,\lambda_{2n}(q),q)|}{\partial b}=
-\frac{\dot{\psi}_+(0)\cdot\psi_{2n}^2(0)}{\psi_+(0)}=(\psi_{2n}\chi_{2n})(0)\,,
$$
where we omit $\lambda\!=\!\lambda_{2n}(q,b)$ and $q$ for short.
\end{proof}

\begin{lemma}\label{*AGradProd} (i) For each $q\in\bH_+$ and $n,m\!\ge\!0$ the following
identities are fulfilled:
$$
\begin{array}{ll}\displaystyle
\left((\psi_{2n+1}^2)',\psi_{2m+1}^2\right)_{+}=0\,,& \displaystyle
\left((\psi_{2n+1}\chi_{2n+1})',\psi_{2m+1}^2\right)_{+}= {\textstyle -
\frac{1}{2}}\,\delta_{mn}\,, \vphantom{\big|_{\big|}}\cr \displaystyle
\left((\psi_{2n+1}^2)',\psi_{2m+1}\chi_{2m+1}\right)_{+}=
{\textstyle\frac{1}{2}}\,\delta_{mn}\,,& \displaystyle
\left((\psi_{2n+1}\chi_{2n+1})',\psi_{2m+1}\chi_{2m+1}
\vphantom{\psi_{2m+1}^2}\right)_{+}=0\,. \vphantom{\big|^{\big|}}\end{array}
$$
(ii) For each $q\in\bH_+$\,, $b\!\in\!\R$ and $n,m\!\ge\!0$ the following identities are
fulfilled:
$$
\left((\psi_{2n}^2)',\psi_{2m}^2\right)_+=
-{\textstyle\frac{1}{2}}\,(\psi_{2n}^2\psi_{2m}^2)(0), \qquad\qquad
\left((\psi_{2n}\chi_{2n})',\psi_{2m}^2\right)_+=
{\textstyle\frac{1}{2}}\,(-\delta_{nm}\!-(\psi_{2n}^2\psi_{2m}\chi_{2m})(0)),
$$
$$
 \left((\psi_{2n}^2)'\!,\psi_{2m}\chi_{2m}\right)_{\!+}\!\!=\!
{\textstyle\frac{1}{2}}\,(\delta_{nm}\!-(\psi_{2n}^2\psi_{2m}\chi_{2m})(0)),  \
\left((\psi_{2n}\chi_{2n})'\!,\psi_{2m}\chi_{2m}\vphantom{\psi_{2m}^2}\right)_{\!+}\!\!=\!
-{\textstyle\frac{1}{2}}\,(\psi_{2n}\chi_{2n}\psi_{2m}\chi_{2m})(0).
$$
\remark {\rm This Lemma is similar to \cite{CKK} Lemma 2.6 (see also \cite{PT} p. 44-45).}
\end{lemma}
\begin{proof}
(i) For instance, we prove the third identity. Integration by parts gives
$$
I_{nm}=\int_{\R_+}(\psi_{2n+1}^2)'(t,q)(\psi_{2m+1}\chi_{2m+1})(t,q)dt=
\frac{1}{2}\int_{\R_+}\{\psi_{2m+1}\chi_{2m+1}\,,(\psi_{2n+1}^2)\}(t,q)dt
$$
$$
=\frac{1}{2}\int_{\R_+}(\chi_{2m+1}\psi_{2n+1}\{\psi_{2m+1}\,,\psi_{2n+1}\}+
\psi_{2m+1}\psi_{2n+1}\{\chi_{2m+1}\,,\psi_{2n+1}\})(t,q)dt\,.
$$
For $n\!\ne\!m$ it follows
$$
I_{nm}=\int_{\R_+} \frac{(\{\psi_{2m+1}\,,\psi_{2n+1}\}\{\chi_{2m+1}\,,\psi_{2n+1}\})'(q,t)}
{2(\l_{2m+1}(q)\!-\!\l_{2n+1}(q))}\,dt=0\,.
$$
If $n\!=\!m$\,, then we have $\{\psi_{2m+1}\,,\psi_{2n+1}\}\!=\!0$ and
$\{\chi_{2m+1}\,,\psi_{2n+1}\}\!=\!1$\,. Therefore,
$$
I_{nn}=\frac{1}{2}\int_{\R_+}\psi_{2n+1}^2(t,q)dt=\frac{1}{2}\,.
$$
The proof of others identities and the proof of (ii) is similar.
\end{proof}

\noindent{\bf A.4 The leading terms of asymptotics of $\psi_+(0,\lambda)$ and
$\psi'_+(0,\lambda)$.} Let
$$
\kappa_n=\p_{+}^0(0,\l_n^0)\,,\qquad \kappa'_n=(\p_{+}^0)'(0,\l_n^0)\,,\qquad
\dot{\kappa}_n=\dot{\p}_{+}^0(0,\l_n^0)\quad {\rm and\ so\ on.}
$$
Note that (\ref{Psi+00lambda}) yields
\begin{equation}\label{KappaValues}
\begin{array}{cc}
\kappa_{2n}\asymp|\lambda_{2n}^0|^{-1/4}\cdot a(\lambda_{2n}^0)\,, \qquad\qquad\quad
\kappa'_{2n}=0\,, & \quad \dot{\kappa}'_{2n}\asymp|\lambda_{2n}^0|^{1/4}\cdot
a(\lambda_{2n}^0)\,, \vphantom{\big|_{\big|}}\cr\vphantom{\big|^{\big|}} \kappa_{2n+1}=0\,,
\qquad \kappa'_{2n+1}\asymp|\lambda_{2n+1}^0|^{1/4}\cdot a(\lambda_{2n+1}^0)\,, & \quad
\dot{\kappa}_{2n+1}\asymp|\lambda_{2n+1}^0|^{-1/4}\cdot a(\lambda_{2n+1}^0)\,,
\end{array}
\end{equation}
and
\begin{equation}\label{KappaSpecAsymp}
\begin{array}{ll}\displaystyle
\frac{\dot{\kappa}_{2n}}{\kappa_{2n}}-\frac{\ddot{\kappa}'_{2n}}{2\dot{\kappa}'_{2n}}=O(n^{-1})\,,
\quad & \displaystyle
\frac{\ddot{\kappa}_{2n}}{\kappa_{2n}}-\frac{\dot{\kappa}_{2n}^2}{\kappa_{2n}^2} +
\frac{\pi^2}{16}=O(n^{-1})\,, \vphantom{\Big|_{\Big|}}\cr \displaystyle
\frac{\dot{\kappa}'_{2n+1}}{\kappa'_{2n+1}}-
\frac{\ddot{\kappa}_{2n+1}}{2\dot{\kappa}_{2n+1}}=O(n^{-1})\,, \quad & \displaystyle
\frac{\ddot{\kappa}'_{2n+1}}{\kappa'_{2n+1}}-
\frac{(\dot{\kappa}'_{2n+1})^2}{(\kappa'_{2n+1})^2} + \frac{\pi^2}{16} =O(n^{-1})\,,
\end{array}\quad n\!\to\!\infty\,.
\end{equation}

\begin{lemma} \label{*LA511} (i) Let $\psi_+^{(1)}\!=\!\psi_+^{(1)}(0,\lambda_{2n}^0,q)$\,,
$\dot{\psi}_+^{(1)}\!=\!\dot{\psi}_+^{(1)}(0,\lambda_{2n}^0,q)$ and so on. The  following
identities and asymptotics are fulfilled for some absolute constant $\delta\!>\!0$:
$$
\begin{array}{ll}
\p_{+}^{(1)}=-\kappa_{2n}\cdot \nc q_{2n}^+\,, &
\dot{\p}_{+}^{(1)}=\kappa_{2n}(-\frac{\dot{\kappa}_{2n}}{\kappa_{2n}}\,\nc q_{2n}^+ +
\frac{\pi^2}{8}\,\nh q_{2n}^+ + \el2_{\frac{1}{4}+\delta}(n))\,, \vphantom{\big|_\big|}\cr
(\p_{+}^{(1)})'=-2\dot{\kappa}'_{2n}\cdot \nh q_{2n}^+\,, &
(\dot{\p}_{+}^{(1)})'=-2\dot{\kappa}'_{2n}(\frac{\dot{\kappa_{2n}}}{\kappa_{2n}}\,\nh q_{2n}^+
+ \frac{1}{2}\,\nc q_{2n}^+ + \el2_{\frac{1}{4}+\delta}(n))\,, \vphantom{\Big|_\big|^\big|}\cr
(\p_{+}^{(2)})'= -2\dot{\kappa}'_{2n}( -\nc q_{2n}^+\nh q_{2n}^+ +
\el2_{\frac{3}{4}+\delta}(n))\,, & \p_{+}^{(2)}=\kappa_{2n}(\frac{1}{2}\,(\nc q_{2n}^+)^2-
\frac{\pi^2}{8}\,(\nh q_{2n}^+)^2 + \el2_{\frac{3}{4}+\delta}(n))\,. \vphantom{\big|^\big|}
\end{array}
$$
(ii) Let $\psi_+^{(1)}\!=\!\psi_+^{(1)}(0,\lambda_{2n+1}^0,q)$\,,
$\dot{\psi}_+^{(1)}\!=\!\dot{\psi}_+^{(1)}(0,\lambda_{2n+1}^0,q)$ and so on. The  following
identities and asymptotics are fulfilled for some absolute constant $\delta\!>\!0$:
$$
\begin{array}{ll}
\p_{+}^{(1)}=-2\dot{\kappa}_{2n+1}\nh q_{2n+1}^+\,, &
\dot{\p}_{+}^{(1)}\!=\!-2\dot{\kappa}_{2n+1}(\frac{\dot{\kappa}'_{2n+1}}{\kappa'_{2n+1}}\,\nh
q_{2n+1}^+ \!+\! \frac{1}{2}\,\nc q_{2n+1}^+ \!+ \el2_{\frac{1}{4}+\delta}(n)),
\vphantom{\big|_\big|}\cr (\p_{+}^{(1)})'=-\kappa'_{2n+1}\cdot \nc q_{2n+1}^+\,,
\vphantom{\Big|_\big|^\big|}&
(\dot{\p}_{+}^{(1)})'\!=\!\kappa'_{2n+1}(-\frac{\dot{\kappa}'_{2n+1}}{\kappa'_{2n+1}}\,\nc
q_{2n+1}^+ \!+\! \frac{\pi^2}{8}\nh q_{2n+1}^+ \!+ \el2_{\frac{1}{4}+\delta}(n)), \cr
\p_{+}^{(2)}\!=\!-2\dot{\kappa}_{2n+1}(-\nc q_{2n+1}^+\nh q_{2n+1}^+ \!+
\el2_{\frac{3}{4}+\delta}(n)), & (\p_{+}^{(2)})' \!=\! \kappa'_{2n+1}(\frac{1}{2}(\nc
q_{2n+1}^+)^2 \!-\! \frac{\pi^2}{8}(\nh q_{2n+1}^+)^2 \!+ \el2_{\frac{3}{4}+\delta}(n)).
\vphantom{\big|^\big|}
\end{array}
$$
\end{lemma}
\begin{proof} See \cite{CKK} Lemmas 5.11, 5.12 and \cite{CKK} Theorem 6.4.
\end{proof}

\vskip 6pt \pagebreak \noindent{\bf A.5 The convergence of one double sum.} Here we prove
technical
\begin{lemma}\label{*ALemma} Let $q,p\!\in\!\bH$\,. Then
$$
S_k= \sum_{n=0}^k\sum_{m=k+1}^{+\infty}
\frac{\left(q,(\psi_{n}\psi_{m})(p)\right)_{L^2(\R)}^2}{m\!-\!n}\to 0\quad as \quad
k\!\to\!\infty\,.
$$
\end{lemma}
\begin{proof}
Using Lemma \ref{*AEstim}, Corollary \ref{*ADotEstim} and asymptotics (\ref{EstimExact}), we
obtain
$$
\psi_n^0=O(\rho^{-1}(x,\lambda_n^0))\,,\qquad \psi_n(x,p)=\psi_n^0(x)+O(n^{-\frac{1}{2}}\log
n\cdot\rho^{-1}(x,\lambda_n^0))\,.
$$
Note that
$$
\biggl(\int_{\R}\frac{|q(t)|dt}{\rho(t,\lambda_n^0)\rho(t,\lambda_m^0)}\biggr)^2\le
\int_\R(t^2\!+\!1)|q(t)|^2dt\cdot
\int_{\R}\frac{dt}{(t^2\!+\!1)\rho^2(t,\lambda_n^0)\rho^2(t,\lambda_m^0)}=
O(n^{-\frac{1}{2}}m^{-\frac{1}{2}})\,.
$$
Let
$$
S_k=\sum_{n=0}^k\sum_{m=k+1}^{k+k^{3/4}}+\sum_{n=0}^k\sum_{m=k+k^{3/4}}^{+\infty}=
S_k^{(1)}+S_k^{(2)}\,.
$$
We have
$$
S_k^{(1)}=O(k^{-\frac{1}{2}})\sum_{n=0}^k
O(n^{-\frac{1}{2}})\sum_{m=k+1}^{k+k^{3/4}}\frac{1}{m\!-\!n}= O(k^{\frac{1}{4}})\sum_{n=0}^k
O(n^{-\frac{1}{2}}(k\!-\!n)^{-1})= O(k^{-\frac{1}{4}}\log k)\,,
$$
i.e. $S_k^{(1)}\!=\!o(1)$ as $k\!\to\!\infty$\,. We estimate $S_k^{(2)}$\,. Let $\ve\!>\!0$ be
sufficiently small. If $m\!\ge\!n$\,, then
$$
\left(q,(\psi_{n}\psi_{m})(p)\right)_{L^2(\R)}^2=
\left(q,(\psi_{n}^0\psi_{m}^0)\right)_{L^2(\R)}^2 + O(\log n\cdot n^{-1}m^{-\frac{1}{2}})
$$
$$
=\biggl(\int_{-n^\ve}^{\,\,n^\ve}\!q(t)\psi_n^0(t)\psi_m^0(t)dt\biggr)^2+
O(n^{-\frac{1}{2}-\frac{\ve}{2}}m^{-\frac{1}{2}})
$$
since
$$
\biggl(\int_{|t|\ge n^\ve}\frac{|q(t)|dt}{\rho(t,\lambda_n^0)\rho(t,\lambda_m^0)}\biggr)^2\le
\int_{|t|\ge n^\ve}(|t|^{3/2}\!+\!1)|q(t)|^2dt\cdot O(n^{-\frac{1}{2}}m^{-\frac{1}{2}})=
O(n^{-\frac{1}{2}-\frac{\ve}{2}}m^{-\frac{1}{2}})\,.
$$
It is easy to see (see e.g. \cite{CKK} Lemma 6.7) that
$$
\psi_n^0(t)= \sqrt{2\big/\pi}\cdot(\lambda_n^0)^{-\frac{1}{4}}\cos(\sqrt{\lambda_n^0}\cdot
t-{\textstyle\frac{\pi n}{2}}) + O(n^{-\frac{3}{4}+3\ve})\,,\qquad
|t|\!\le\!n^{\varepsilon}\,.
$$
If $m\!\ge\!n\!+\!n^{3/4}$\,, then $\sqrt{\lambda_m^0}\!-\!\sqrt{\lambda_n^0}\!\ge
n^{\frac{1}{4}}$\,. Integration by parts and $q'\!\in\!L^2(\R)$ follow
$$
\int_{-n^\ve}^{\,\,n^\ve}\!q(t)\psi_n^0(t)\psi_m^0(t)dt=
O(n^{-\frac{1}{2}}m^{-\frac{1}{4}})\cdot \int_{-n^\ve}^{\,\,n^\ve}\!|q'(t)|dt +
O(n^{-\frac{3}{4}+3\ve}m^{-\frac{1}{4}})= O(n^{-\frac{1}{2}+\frac{\ve}{2}}m^{-\frac{1}{4}})
$$
Summarizing, we obtain $\left(q,(\psi_{n}\psi_{m})(p)\right)_{L^2(\R)}^2\!=\!
O(n^{-\frac{1}{2}-\frac{\ve}{2}}m^{-\frac{1}{2}})$\,, if $m\!\ge\!n\!+\!n^{3/4}$\,. Therefore,
$$
S_k^{(2)}=\sum_{n=0}^k\sum_{m=k+k^{3/4}}^{+\infty}
\frac{O(n^{-\frac{1}{2}-\frac{\ve}{2}}m^{-\frac{1}{2}})}{m\!-\!n}\le \sum_{n=0}^k
O(n^{-\frac{1}{2}-\frac{\ve}{2}})\sum_{m=k+1}^{+\infty}\frac{O(m^{-\frac{1}{2}})}{m\!-\!n}
$$
Note that $\sum_{n=0}^k
O(n^{-\frac{1}{2}-\frac{\ve}{2}})\!=\!O(k^{\frac{1}{2}-\frac{\ve}{2}})$ and
$\sum_{m=k+1}^{+\infty}\frac{O(m^{-\frac{1}{2}})}{m\!-\!n}\!=\!O(\log k\cdot
k^{-\frac{1}{2}})$\,. Hence, $S_k^{(2)}\!=\!o(1)$ as $k\!\to\!\infty$\,. We are done.
\end{proof}

\end{document}